\documentclass[lettersize,journal]{IEEEtran}
\usepackage{amsmath,amsfonts}
\usepackage{algorithmic}
\usepackage{algorithm}
\usepackage{array}
\usepackage[caption=false,font=normalsize,labelfont=sf,textfont=sf]{subfig}
\usepackage{textcomp}
\usepackage{stfloats}
\usepackage{url}
\usepackage{verbatim}
\usepackage{graphicx}
\usepackage{cite}
\usepackage[colorlinks=true,
            linkcolor=blue,   
            citecolor=blue,  
            urlcolor=black]{hyperref}
\hypersetup{pdfborder={0 0 0}}
\usepackage[normalem]{ulem} %
\usepackage{soul} %

\usepackage{xcolor} %

\begin{document}

\title{Closed-Loop Transmission Power Control for Reliable and Low-Power BLE Communication in Dynamic IoT Settings}

\author{Ziyao Zhou,~\IEEEmembership{}
        Hen-Wei Huang\,~\IEEEmembership{}%
\thanks{This work was supported by the Nanyang Professorship, the MOE Tier 1 grant RG71/24, and the MTC MedTech Programmatic Fund M24N9b0125. \textit{(Corresponding author: Hen-Wei Huang.)}}%
\thanks{Ziyao Zhou and Hen-Wei Huang are with the School of Electrical and Electronic Engineering, 
Nanyang Technological University, Singapore (e-mail: ZHOU0557@e.ntu.edu.sg; henwei.huang@ntu.edu.sg). Hen-Wei Huang is also with the LKC School of Medicine, Nanyang Technological University, Singapore.}%
}

\markboth{IEEE Internet of Things Journal,~Vol.~XX, No.~X, Month~Year}%
{Shell \MakeLowercase{\textit{et al.}}: A Sample Article Using IEEEtran.cls for IEEE Journals1}


\maketitle

\begin{abstract}
Reliable and energy-efficient Bluetooth Low Energy (BLE) communication is crucial for Internet of Things (IoT) applications in dynamic environments. However, the Received Signal Strength Indicator (RSSI) and data throughput in BLE are highly susceptible to environmental variability, affecting communication performance. In this work, we systematically analyze the interdependence between RSSI, throughput, transmission power (TXP), and the peripheral device’s system power consumption under various real-world conditions, observing that adjusting the TXP effectively influences both the RSSI and throughput. We propose a robust closed-loop TXP control framework based on Proportional-Integral-Derivative (PID) controllers. Two initial control strategies were implemented: an RSSI-based approach and a throughput-based approach, each with distinct advantages and limitations. The RSSI-based method offers rapid responsiveness to signal fluctuations but lacks direct correlation with data throughput, while the throughput-based method provides more accurate feedback on transmitting throughput at the cost of slower response. To address these limitations, a hybrid RSSI–throughput control strategy was developed, combining the responsiveness of RSSI feedback with the accuracy of throughput measurements. This hybrid approach effectively maintains data throughput near the target level with minimal variance, even under rapidly changing environmental conditions.

\end{abstract}


\section{Introduction}

Bluetooth Low Energy (BLE) technology has become increasingly prevalent in the Internet of Things (IoT) field \cite{Gautam2024LowPower,11213141, s25040996, 2423425, 8832176, 10659117 }, driven by its low power consumption and suitability for multiple communication applications. It has a maximum data throughput of 2 Mbps, and power consumption is about 10 mW \cite{1}.

\begin{figure}[htb]
 \centering
\includegraphics[width=1\linewidth]{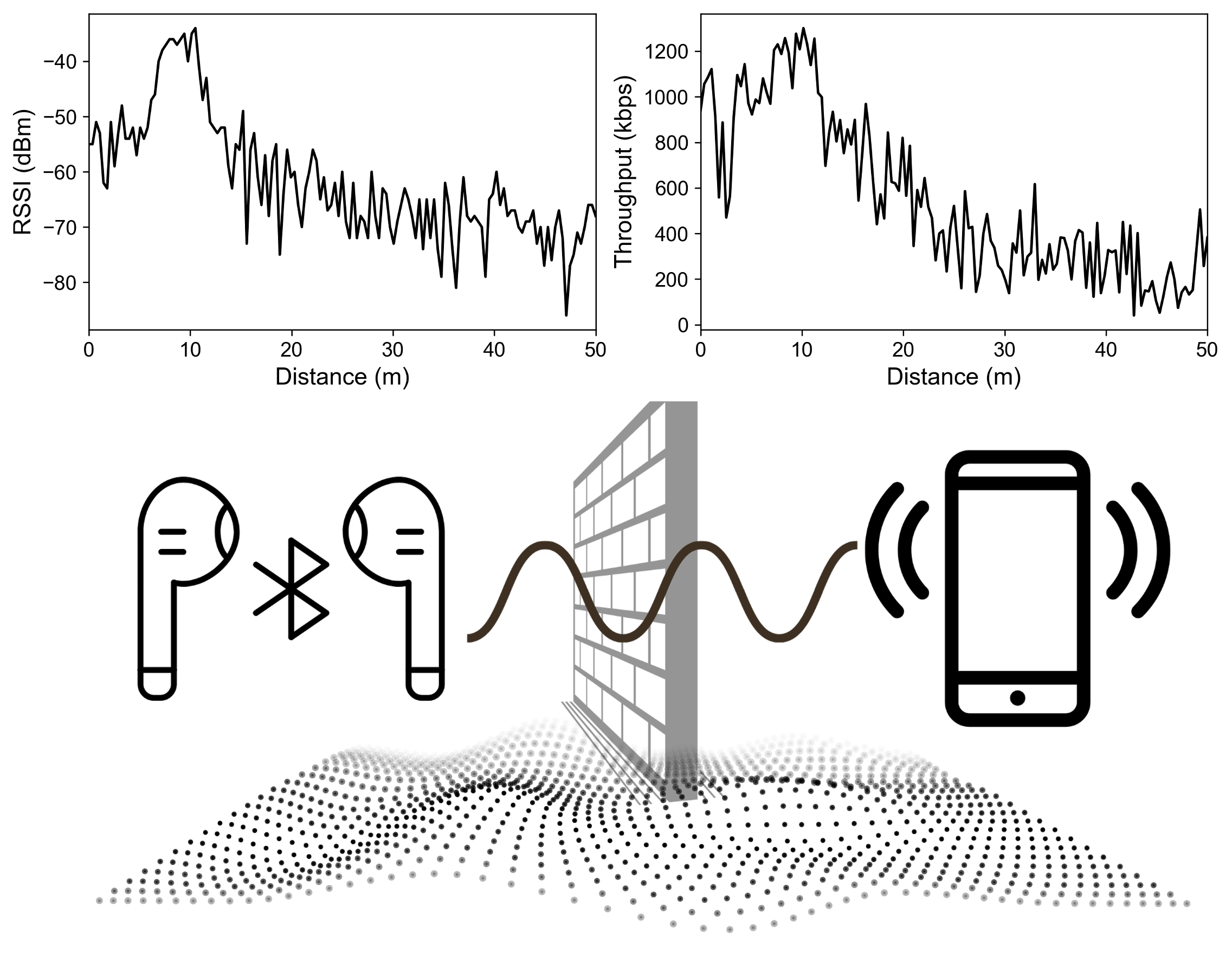}
\caption{RSSI and throughput variation versus distance for BLE communications.}
\label{fig Throughput&RSSI variation}
\end{figure}

One major challenge for wireless communication, including BLE, is that the Received Signal Strength Indicator (RSSI) \cite{7477344} and data throughput \cite {pang2024modelingtradeoffthroughputreliability} can easily be affected by ambient environments. Both parameters’ variation in longer distances shows a decreasing trend, as shown in Figure~\ref{fig Throughput&RSSI variation}. RSSI variations in dynamic environments are influenced by three components: shadowing, fading, and path loss \cite{5349183}. The throughput is influenced by three elements: path loss, noise floor, and channel effect \cite{4526126}. These factors introduce disturbances, making communication inherently unstable.

To ensure a high quality of service, the BLE system’s RSSI should be higher than the specific value to avoid disconnection \cite{Ramirez2021A}. In addition, the system’s throughput should also exceed the specific demands of the target application, thereby guaranteeing stable communication performance. For example, BLE Audio implementations employing the low complexity communication codec (LC3)  require data rates up to 320 kbps with 10-millisecond frame intervals \cite{BluetoothLC3}, while human motion tracking typically demands approximately 80 kbps when sampling at 100 Hz \cite{8501957}. Generally, the user selects the highest transmission power (TXP) to overcome RSSI/throughput instability \cite{qureshi2019evaluating}. However, these methods significantly compromise the energy efficiency.

Several algorithms have been proposed to maintain communication stability while improving energy efficiency. Adaptive Transmission Power Control (ATPC), proposed in \cite{10.1145/2746342}, utilizes the RSSI and Link Quality Indicator (LQI) feedback to optimize energy efficiency while maintaining link quality. ATPC employs a closed-loop control mechanism for dynamic TXP adjustment. However, its reliance on LQI, defined in IEEE 802.15.4 \cite{10794632}, restricts its applicability to BLE-based communication systems, where LQI is unavailable. Furthermore, ATPC's performance in rapidly changing wireless environments remains unexplored. These limitations underscore the need for alternative control strategies tailored to BLE and capable of robust operation under dynamic conditions.


To improve BLE application efficiency under varying wireless conditions, Bluetooth v5.2 introduced the Low Energy Power Control (LEPC) mechanism \cite{bluetooth2025core52}. LEPC is intended to maintain the receiver's signal-to-noise ratio (SNR) within an optimal operating range, often referred to as the “golden range.” While the adaptation relies on accurate RSSI measurements, the Bluetooth specification leaves the exact algorithm design and the definition of this target range to the discretion of the developer \cite{10770502}.

Another relevant one is AdaptaBLE, a BLE-specific solution introduced in \cite{9013624}, which regulates TXP based on the Packet Reception Ratio (PRR). Its reliance on a non-robust conditional judgment control algorithm limits its adaptability to dynamic environments, and the PRR metric, which is derived over extended periods \cite{10.1145/2746342}, constrains real-time responsiveness.

The most advanced method is described in \cite{10770502}, making use of both RSSI and PRR information in BLE to tune the TXP. Though RSSI is a fast-updating parameter that can partially compensate for the latency of PRR, PRR indirectly influences throughput, as throughput is also related to packet length and packet rate. And it employs non-robust BLE TXP control strategies based on conditional judgment.

Existing BLE-oriented methods still leave a gap for improvement from a control perspective. First, they generally neglect the impact of communication latency, which introduces time delays that are critical for the stability and responsiveness of control. Second, most approaches rely on a single parameter and a simple control rule, without considering the diverse dynamic characteristics of different feedback signals. In emerging IoT scenarios, both communication latency and control delay must be carefully considered \cite{10107704, DEEP2025125251, 10717993}. Moreover, dual-loop control architecture using different feedback parameters has demonstrated enhanced control capability and improved system robustness under such conditions\cite{10.1007/978-981-19-2188-9_4}.

Furthermore, existing BLE methods regulate throughput indirectly using RSSI and PRR. Although throughput is the most intuitive communication metric, no study has directly utilized calculated throughputs for TXP control.

These observations motivate the need for a new BLE transmission power control framework. In this study, we propose a closed-loop TXP control framework based on Proportional-Integral-Derivative (PID) controllers implemented on the central device. The controller dynamically adjusts the TXP of the peripheral device using real-time RSSI \cite{zhou2025adaptive} or throughput feedback \cite{zhou2025throughput} collected at the central device. The final strategy integrates both RSSI and throughput to exploit their complementary advantages—combining the rapid responsiveness of RSSI with the intuitive accuracy of throughput—to achieve robust, adaptive TXP regulation. This unified approach addresses the limitations observed in prior individual strategies and significantly improves communication stability.


This study first investigates the interrelationship among throughput, RSSI, TXP, and system power consumption under various environmental conditions. Subsequently, the effectiveness of the RSSI-based and throughput-based control methods is independently validated. Finally, we integrate both approaches into a unified RSSI+throughput control scheme, demonstrating that our combined method significantly enhances both communication stability and energy efficiency across all evaluated scenarios.

The main contributions of this work are summarized as follows:

\begin{itemize}
    \item A comprehensive analysis was conducted on the interdependence among RSSI, throughput, TXP, and power consumption in dynamic BLE IoT environments, explicitly considering the effects of communication latency and control delay.
    
    \item Two PID-based closed-loop strategies—RSSI-based and throughput-based—were designed and comparatively evaluated, highlighting their respective responsiveness and accuracy characteristics.
    
    \item A dual-loop hybrid RSSI–throughput control framework was proposed and experimentally validated, demonstrating improved throughput stability, reduced variance, and enhanced energy efficiency.
\end{itemize}

\section{Methodology for Investigating the Interrelationship Among Throughput, RSSI, TXP, and System Power Consumption}

In this study, we utilized a pair of combined development platforms consisting of the Nordic nRF54L15 DK and the nRF21540 front-end module (FEM). One combined board was configured as the central device (receiver), responsible for acquiring real-time RSSI and throughput values, while concurrently receiving application data. The other combined board served as the peripheral device (transmitter), typically low-power, resource-constrained devices that provide data, continuously transmitting a fixed 244-byte payload. Data communication between the two devices was conducted via the Nordic UART Service.

The experiments were conducted under the following BLE communication parameters:
\begin{itemize}
  \item ATT Maximum Transmission Unit (MTU) size: 498 bytes
  \item Connection interval: 320 units (equivalent to 400 ms)
  \item Physical layer (PHY) data rate: 2 Mbps
\end{itemize}

\subsection{Impact of nRF21540 FEM on System Power Consumption}

To increase the upper limit of TXP, we integrated the nRF21540 FEM with the nRF54L15 DK. The addition of the nRF21540 FEM enables the TXP to be raised from a maximum of 8 dBm to 20 dBm \cite{nRF21540Spec}. For most IoT applications, higher TXP ensures greater communication range and improved link stability \cite{Fahrezi2024Analisis}. However, this enhancement comes at the cost of increased overall system power consumption \cite{Li2018Reducing}.

To quantitatively assess this trade-off, we used the Nordic Power Profiler Kit II (PPK) as the ampere meter to monitor the peripheral device, which enables high-resolution current measurements at a sampling rate of 100,000 samples per second \cite{nordicsemi_ppk2_userguide_v1.0.1}. The nRF54L15 DK is powered by 1.8 V, allowing the total power consumption to be calculated from the current. We compared the energy consumption of the peripheral device under two configurations: with the  FEM (TXP = 20 dBm) and without it (TXP = 8 dBm).

Since transmission throughput also affects power consumption, we additionally evaluated system-level energy usage under three distinct throughput levels: approximately 2 kbps, 600 kbps, and the maximum achievable 1300 kbps. These throughput levels were configured by adjusting the timer interval that controls the packet transmission frequency. In this experiment, the peripheral and central devices were placed in direct contact with zero separation in the lab to eliminate the influence of signal attenuation.

\subsection{Impact of Calculation Frequency on RSSI and Throughput}
        
RSSI represents the signal strength measured at the receiver side \cite{Mohsin2019Study}, and can be directly retrieved from the hardware register of the nRF54L15 chip. This chip updates RSSI values within 20 microseconds upon signal level changes \cite{nRF54LSeriesSpec}. In contrast, throughput reflects the amount of data successfully transmitted over a connection within a given time interval \cite{1047314}, which must be computed at the receiver side over a defined period. This experiment aims to investigate how different calculation frequencies affect the obtained values of both RSSI and throughput. 

The experimental setup was established in the indoor lab, with the central device and the peripheral device separated by a distance of 5 m. The peripheral device acted as the transmitter, sending data at the maximum throughput with its TXP fixed at 20 dBm. The central device measured RSSI values and computed the data throughput at calculation frequencies ranging from 0.001 Hz to 1000 Hz, using six intervals with seven points. The collected results were statistically analyzed, with the mean, median, and standard deviation of each calculation frequency setting used to assess the stability and accuracy of the measurements.

\subsection{Relationship Between RSSI and Throughput in Different Environments}

Based on the previous experiment, we selected an appropriate calculation frequency for throughput measurement. The TXP of the peripheral device was fixed at 0 dBm, and experiments were carried out in three distinct environments: the rooftop, the corridor, and the indoor laboratory, as environmental factors affect throughput performance \cite{8419192}. The rooftop offered the cleanest environment, characterized by open space and minimal interference, making it the ideal reference setting for performance evaluation. The corridor, although free from active electronic devices, presented moderate complexity due to multipath reflections and scattering along the walls. In contrast, the indoor lab represented the most challenging scenario for BLE communication, with a high density of electronic equipment and furniture causing substantial electromagnetic interference and signal attenuation.

During the experiment, the peripheral device was kept stationary, while the central device was mounted on a movable platform to allow controlled variation in the distance between them. The platform was controlled to move and cover the range of RSSI values from –20 dBm to –80 dBm three times. At each RSSI level, the corresponding throughput was measured multiple times to capture its variation and establish a relationship between RSSI and throughput in different environments.

\subsection{Resolution of TXP Adjustment}
The nRF54L15 used in this study supports a configurable output power of up to +8 dBm, with fine-grained resolution steps of 1 dB from –10 dBm to +8 dBm \cite{nRF54LSeriesSpec}. The available TXP settings and their corresponding resolution are summarized in Table~\ref{tab:txpower_resolution}.

\begin{table}[t]
    \centering
    \caption{TXP Setting Resolution for nRF54L15}
    \renewcommand{\arraystretch}{1.1}
    \begin{tabular}{|c|c|c|}
        \hline
        \textbf{Output‐Power Range (dBm)} & \textbf{Step Size (dB)} & \textbf{Number of Steps} \\
        \hline
        +8 to –10    & 1   & 18 \\
        –10 to –22   & 2   & 6  \\
        –22 to –28   & 6   & 1  \\
        –28 to –40   & 12  & 1  \\
        –40 to –46   & 6   & 1  \\
        \hline
    \end{tabular}
    \label{tab:txpower_resolution}
\end{table}

After integrating the nRF21540 FEM, the actual output TXP of the device becomes uncertain due to additional amplification stages. To characterize this, we conducted a systematic sweep of TXP requests from -50 dBm to +50 dBm on the peripheral device. The values returned by the chip were logged for each request. Simultaneously, an RF power meter was connected to the output port to validate whether the actual transmitted power matched the reported software-configured values.

\subsection{Impact of TXP Adjustment on RSSI, Throughput and System Power Consumption Across Three Environments}

In the three environments—rooftop, corridor, and lab—the peripheral and central devices were positioned 5 m apart. Based on the previous characterization, we selected six valid TXP settings that could be reliably configured. For each setting, the peripheral side was connected to the PPK to record power consumption, while on the central side, the corresponding RSSI and throughput values were recorded in all three environments to evaluate the impact of TXP adjustment on communication performance under different environmental conditions.

\subsection{Relationship Between Throughput and Overall System Power Consumption}

In the previous TXP adjustment experiment, the PPK was used to record the power consumption of the peripheral device. Simultaneously, throughput data from the central device was also collected. To investigate the correlation between central throughput and peripheral system power consumption, we selected the lab environment as a representative example.

As the signal strength becomes lower, the probability of packet loss also rises \cite{De2017Finding}. Frequent disconnections during connection events may lead to a reduction in overall power consumption.

\subsection{Latency from Central to Peripheral Device}
To enable TXP updates based on feedback from the central device, it is essential to quantify the latency involved in updating the peripheral’s TXP. For this purpose, we configured the central device to toggle a designated General-Purpose Input/Output (GPIO) pin: the pin is pulled high for 10 ms immediately after a data packet is transmitted, then pulled low. Similarly, the peripheral device toggles its own GPIO pin upon successfully receiving the data packet. Both GPIO signals were monitored using two channels of a single oscilloscope, allowing us to determine the latency by measuring the time difference between the rising edges of the corresponding waveforms. This procedure was repeated across ten independent trials, and the average latency was reported.

\section{Methodology for Deploying the PID Controller}

\begin{figure}[t]
 \centering
\includegraphics[width=0.9\linewidth]{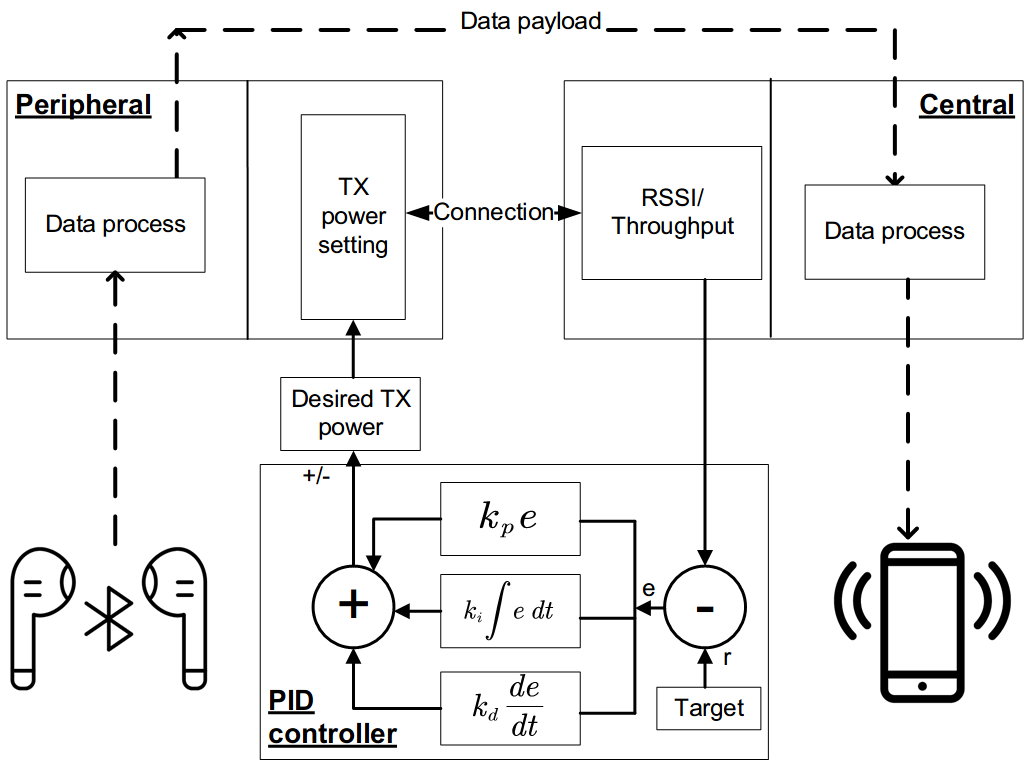}
\caption{Block diagram of the RSSI/throughput PID controller for TXP regulation.}
\label{fig:RSSI_Throughput_PID_Block}
\end{figure}

\subsection{RSSI-based Methodology}

Our PID system setup consists of two traces, as illustrated in Fig.~\ref{fig:RSSI_Throughput_PID_Block}. The first trace, indicated by dashed lines, represents the data flow of the BLE application, encompassing data generation at the peripheral, transmission over the BLE link, and subsequent processing at the central device. The second trace, shown with solid lines, implements a closed-loop PID controller. The central device continuously measures the RSSI and feeds the value into the PID controller. Based on the difference between the target RSSI and the measured RSSI, the controller computes the appropriate TXP for the peripheral. All control logic and computation are executed on the central device to minimize the computational and power burden on the peripheral.

To better understand the additional power consumption introduced by implementing PID control, we evaluated its impact on both the central and peripheral devices. Specifically, we varied the PID computation frequency from 0.01 Hz to 1000 Hz and measured the resulting power overhead. On the central side, we quantified the power consumed for calculating the target TXP and transmitting it to the peripheral. On the peripheral side, we measured the power required to receive and apply the updated TXP. This analysis provides insight into the cost of real-time control under different update rates.

We selected an appropriate frequency for PID computation and TXP updates based on the power consumption and frequency responsiveness. For the RSSI-based method, the experiment was conducted on the rooftop, where the electromagnetic environment was the least complex. The RSSI target is chosen to be -60 dBm.

For the PID controller, through iterative adjustment, the optimal parameters were determined as $k_p = 0.2$ and $k_i = 0.01$. To prevent excessive overshoot and facilitate rapid convergence, the PID output was constrained within a limit of 2.

During the test, the peripheral device remained stationary while the central device moved at a constant speed, gradually increasing the distance between the two devices from 0 to 50 m. Three strategies are compared: the RSSI-based method, fixed TXP of 20 dBm, and -10 dBm, respectively. Each strategy is conducted three times. The PPK was used to record the power consumption of the peripheral device. Meanwhile, the RSSI values measured by the central device, along with the TXP values determined by the controller, were also recorded for analysis.

Subsequently, in the lab environment with a fixed distance of 30 cm between the peripheral and central devices, we evaluated the performance of the RSSI-based method under extreme conditions by setting the target RSSI to –65 dBm. The FEM provides a receive gain of 13 dB \cite{nRF21540Spec}; within this range, removing the FEM causes a sharp decline in RSSI, bringing it close to the receiver sensitivity threshold of the 2 Mbps PHY layer (around -90 dBm for nRF54L15), which may result in link disconnection. By physically unplugging the FEM under these controlled conditions, we aim to assess whether the RSSI-based control strategy can maintain stable communication despite abrupt degradation in signal strength.

\subsection{Througput-based Methodology}
Similar to the RSSI-based approach, the throughput-based method employs the same controller design, as illustrated in Fig.~\ref{fig:RSSI_Throughput_PID_Block}. In this setup, the central device performs computations based on the measured throughput and transmits the calculated desired TXP to the peripheral device. To quantify the power overhead introduced by this method, we measured the additional power consumption on the central device for computing and transmitting the desired TXP, and on the peripheral device for applying the TXP update, across a range of controller update frequencies from 0.01 Hz to 1000 Hz.

We applied a similar methodology to evaluate the throughput-based control approach. The PID is set to the optimal calculation frequency. The PID values were selected explicitly as $k_p = 0.009$ and $k_d = 0.0001$ with a target of 800 kbps. Additionally, we set the PID output limit to 2, ensuring that the system quickly converges to the target value without excessive overshoot. 


The peripheral device’s power consumption was monitored using the PPK, while the central device recorded both the throughput and the TXP values determined by the controller. During the experiment, the peripheral and central devices were placed in a corridor environment and moved apart at a constant speed, covering a distance from 0 to 50 m. In this environment, three strategies are compared: the throughput-based method, TXP of 20 dBm, and TXP of -10 dBm, respectively. Each strategy is conducted three times. 

Subsequently, in the lab setting with a fixed distance of 30 cm between the peripheral and central devices, the throughput target was set to 100 kbps. In this throughput range, the corresponding RSSI typically falls around -65 dBm. Removing the FEM under these conditions causes the RSSI to approach the receiver sensitivity threshold. By physically disconnecting the FEM, we aim to assess whether the throughput-based control method can effectively compensate for such abrupt changes and maintain communication stability.

\subsection{RSSI+Througput Methodology}
\begin{figure}[t]
 \centering
\includegraphics[width=1\linewidth]{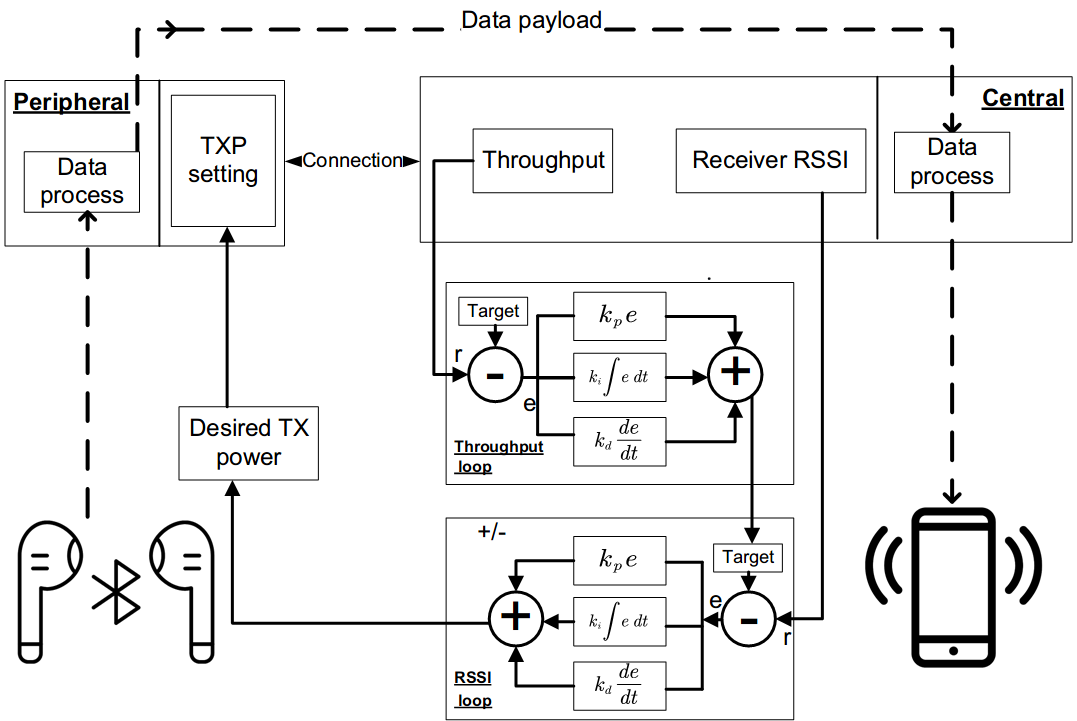}
\caption{Block diagram of the RSSI+Throughput PID controller for TXP regulation.}
\label{fig:RSSI+Throughput_PID_Block}
\end{figure}

To simultaneously incorporate RSSI and throughput, we designed a cascaded PID controller as illustrated in Fig.~\ref{fig:RSSI+Throughput_PID_Block}. In this architecture, the throughput loop computes the RSSI target at an optimal frequency based on the throughput-based method, while the RSSI loop updates the TXP at another optimal frequency based on the RSSI method. The PID parameters for the RSSI loop are: $k_p = 0.009$, $k_i = 0.00$, $k_d = 0.0001$; for the throughput loop are: $k_p = 0.1$, $k_i = 0.00$, $k_d = 0.01$. Similarly, we used a PPK to record the power consumption of PID computation + transmission on the central device and TXP updates on the peripheral device. 

Subsequently, we set the throughput target to 800 kbps and initialized the RSSI target at -60 dBm. An experiment was conducted in the most complex indoor lab environment, where the peripheral and central devices were gradually separated from 0 to 50 m at a constant speed. In this environment, the performance of the combined RSSI+throughput method is compared against the previously proposed individual RSSI-based and throughput-based approaches. Each method is conducted three times.

During this process, the peripheral system’s power consumption, the central device’s RSSI target, measured RSSI, throughput, and the controller-selected TXP of the cascaded PID controller were all recorded.

Finally, in the lab setting, we positioned the peripheral and receiver 30 cm apart and adjusted the throughput target to 100 kbps. Under these conditions, we were able to reach the critical operating point using the FEM. By manually detaching the FEM, we further evaluated whether the proposed method could maintain stable communication under disturbance.




\section{Results on the Interrelationship Among Throughput, RSSI, TXP, and System Power Consumption}

\subsection{Impact of nRF21540 FEM on System Power Consumption}
\begin{figure}[t]
 \centering
\includegraphics[width=1\linewidth]{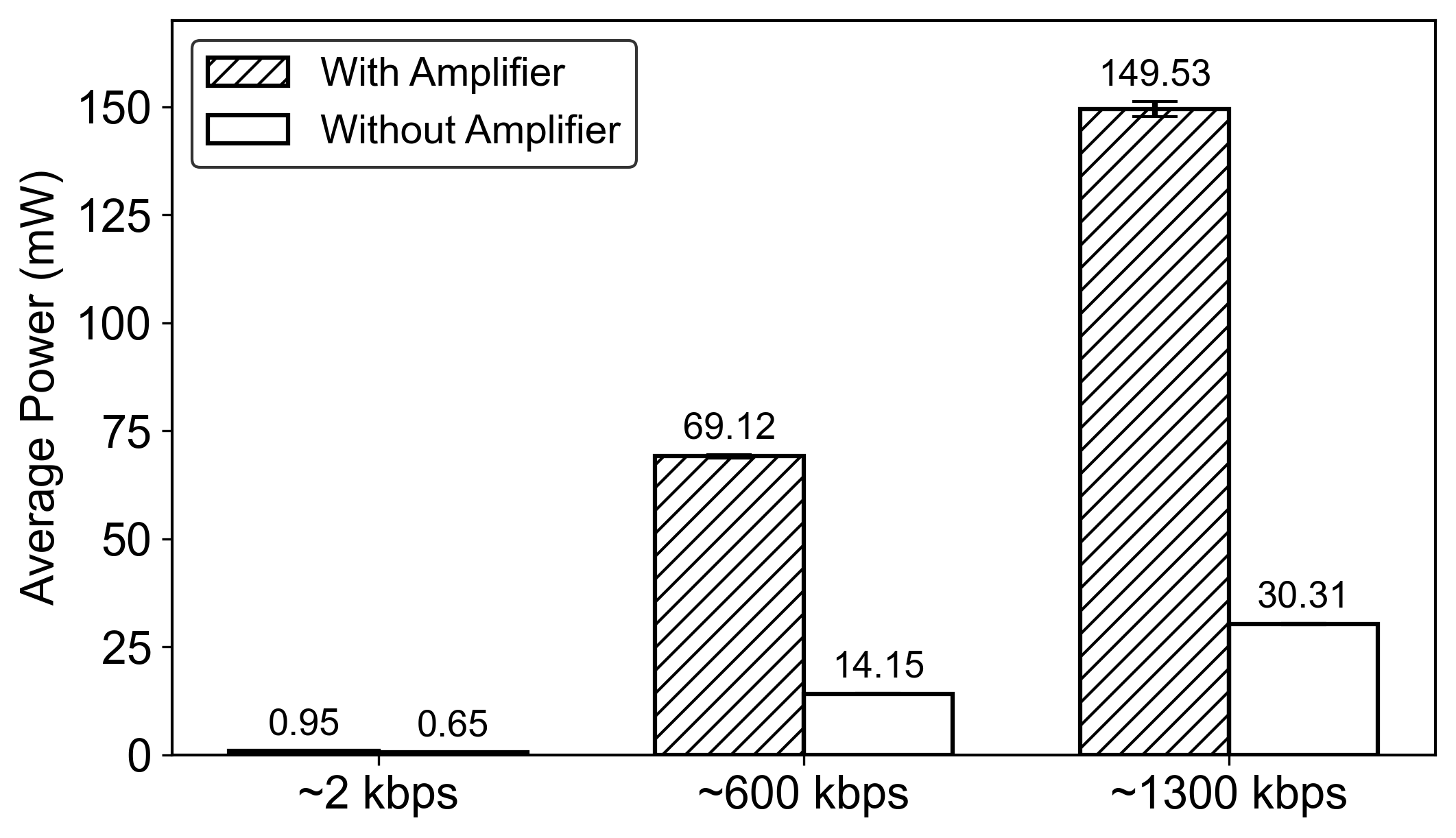}
\caption{Impact of the nRF21540 FEM on system power consumption under different throughput levels.}
\label{fig:FEM_power_impact}
\end{figure}

We employed the nRF21540 FEM to boost the TXP from 8 dBm to 20 dBm. As shown in Fig.~\ref{fig:FEM_power_impact}, the system power consumption with and without the FEM was measured under three different throughput conditions.

Higher data throughput corresponds to higher overall system power consumption. Specifically, without the use of a FEM, the power consumption is approximately 0.65 mW at low throughput, 14.15 mW at medium throughput, and 30.31 mW at high throughput.

At low throughput levels ($\sim2\text{kbps}$), the power consumption remains nearly unchanged regardless of whether FEM is integrated. However, at higher throughput levels, incorporating the FEM results in a substantial increase in power consumption. Specifically, at $\sim600\text{kbps}$ and $\sim1300\text{kbps}$, the power consumption increases by approximately fivefold. In conclusion, while the inclusion of the FEM significantly enhances the maximum achievable TXP range, it also leads to a marked increase in overall power consumption.

\subsection{Impact of Calculation Frequency on RSSI and Throughput}
\begin{figure}[t]
 \centering
\includegraphics[width=1\linewidth]{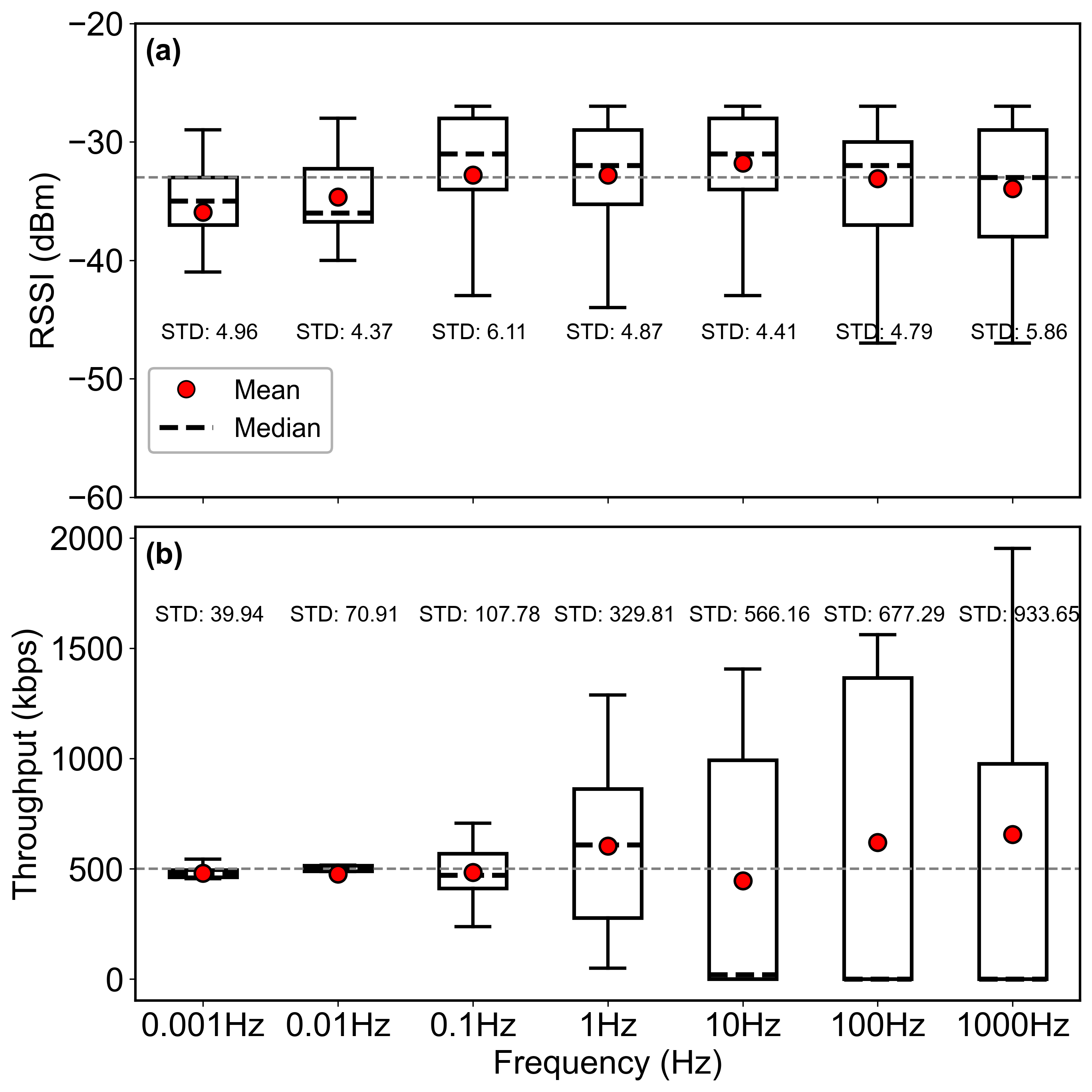}
\caption{The impact of calculation frequency on the measured results at a fixed distance. (a) RSSI results; (b) throughput results. STD: standard deviation.}
\label{fig:sampling_freq_impact}
\end{figure}

In the previous experiment, adding FEM significantly enhances the maximum achievable TXP range. Therefore, we configured both the peripheral and central devices with the nRF54L15 DK + nRF21540 FEM, setting the peripheral’s TXP to 20 dBm and placing it at a distance of 5 m from the central device in a lab. RSSI and throughput were measured using different calculation frequencies, and the corresponding results are shown in Fig.~\ref{fig:sampling_freq_impact}.

Fig.~\ref{fig:sampling_freq_impact}(a) shows the distribution of RSSI values obtained at different calculation frequencies. In each box, the red dot represents the mean, and the dashed horizontal line indicates the median. The standard deviation (STD) is also annotated in the figure. As observed, both the mean, median, and STD of RSSI remain approximately the same across all frequencies. The mean and median RSSI values were centered around -33 dBm, with a standard deviation of approximately 5dB, indicating that the calculation rate has minimal impact on RSSI measurements. This consistency arises from the fact that the nRF54L15 chipset updates RSSI every 20 µs, which is significantly faster than any of the selected calculation intervals, thereby ensuring high temporal resolution and reliable real-time performance.


Fig. \ref{fig:sampling_freq_impact}(b) shows that the mean and median throughput values remain relatively stable across different calculation frequencies; however, the STD increases significantly as the calculation frequency rises. This growing variability arises from the discrete, packet-based nature of BLE communication. At higher calculation frequencies, each monitoring interval becomes so short that it may fail to capture any complete packet transmissions, resulting in throughput values close to zero. Conversely, some intervals may coincidentally contain multiple packet arrivals, leading to abnormally high throughput readings. These random misalignments between the calculation intervals and packet arrival times introduce substantial fluctuations, thereby broadening the distribution of measured throughput values and increasing the observed STD.

For subsequent calculations, we selected a calculation frequency of 1 Hz for both RSSI and throughput.
 
\subsection{Relationship Between RSSI and Throughput in Different Environments}

\begin{figure}[t]
 \centering
\includegraphics[width=1\linewidth]{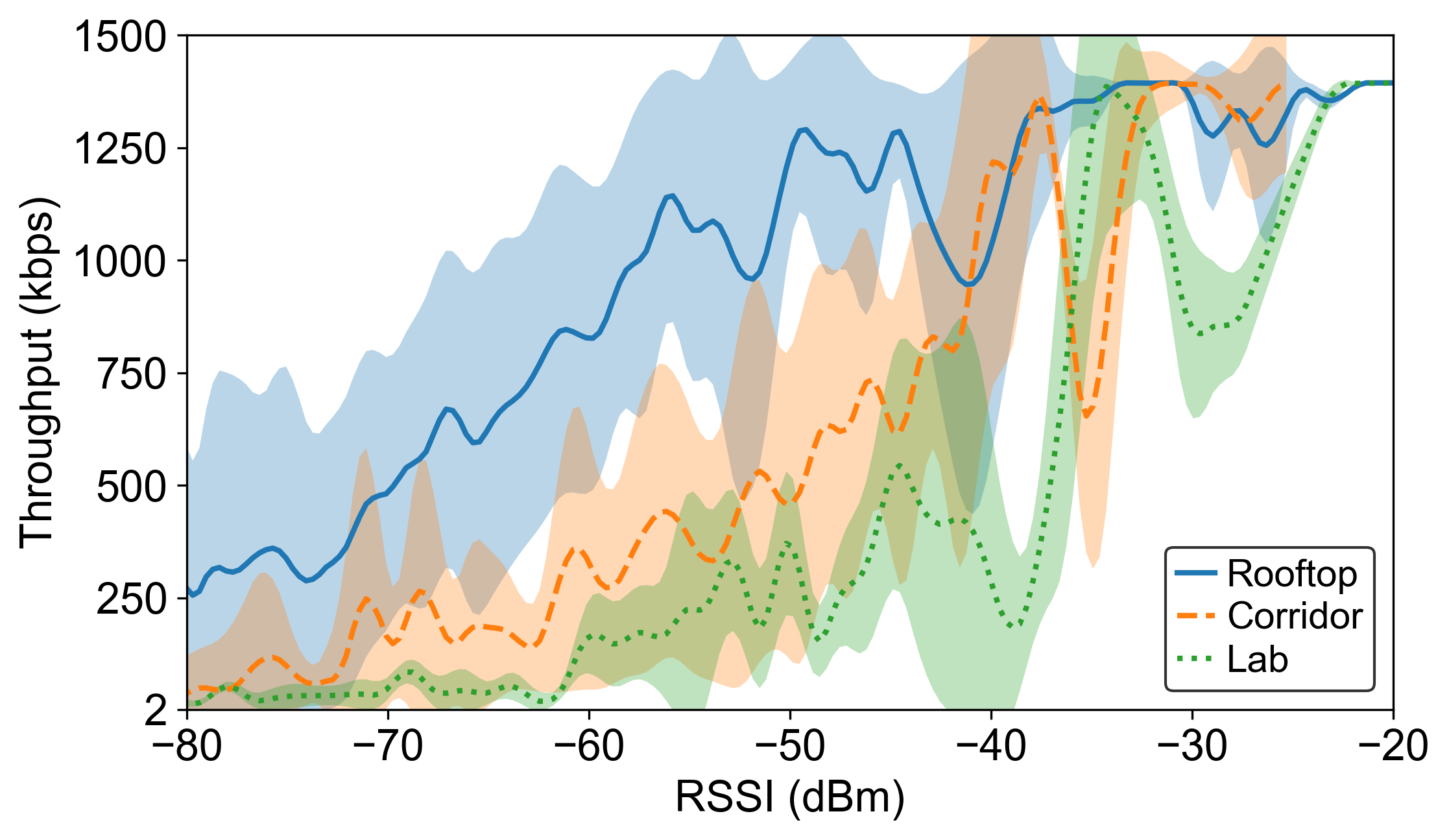}
\caption{Relationship between RSSI and throughput across different environments.}
\label{fig:rssi_throughput_env}
\end{figure}

We evaluated the relationship between RSSI and throughput under three different environmental conditions, as shown in Fig.~\ref{fig:rssi_throughput_env}. Overall, throughput exhibits a positive correlation with RSSI: higher RSSI values generally correspond to higher throughput. In all three environments, as long as the RSSI is sufficiently high, the throughput can reach the maximum value of 1300 kbps. Notably, even at low RSSI levels around –80 dBm, all three curves maintained a minimum throughput of approximately 2 kbps, provided the connection remained active.

However, when comparing results across environments, it is evident that environmental factors affect throughput differently. Specifically, in the rooftop scenario, which represents a relatively clean electromagnetic environment, the throughput remained above 1000 kbps across a wide RSSI range from -60 dBm to -20 dBm. In contrast, the other two environments required significantly higher RSSI levels to sustain throughput above 1000 kbps.

\subsection{Resolution of TXP Adjustment}

\begin{figure}[t]
 \centering
\includegraphics[width=1\linewidth]{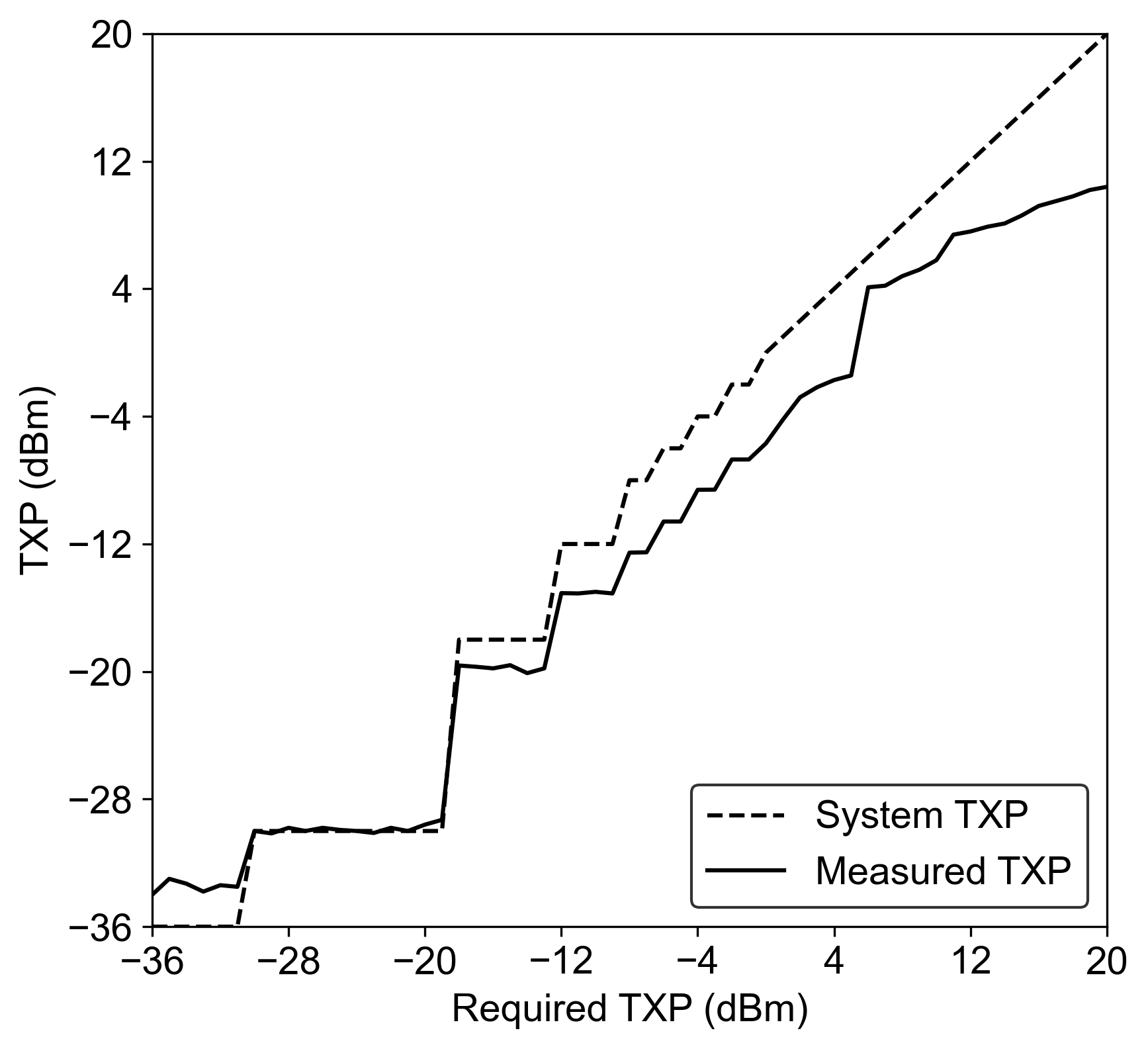}
\caption{Resolution and accuracy of TXP adjustment: comparison between requested and measured TXP levels.}
\label{fig:txp_resolution}
\end{figure}

We cyclically requested TXP values ranging from -50 dBm to 50 dBm from the chip and observed that it reliably responded within the range of -36 dBm to 20 dBm. Within this functional range, we recorded both the TXP values reported by the chip and the actual TXP measured using an RF power meter. The results are shown in Fig.~\ref{fig:txp_resolution}.

It can be observed that the granularity of TXP adjustment is low in the lower TXP range: there is only one adjustable TXP value between -36 and -31 dBm, as well as between -30 and -19 dBm. In contrast, the higher TXP range, from 0 to 20 dBm, supports 1 dB steps, allowing for finer adjustment precision. Moreover, although the TXP values measured by an external RF meter generally align with those reported by the chip, noticeable discrepancies are observed in the high TXP region. The experimental results suggest that adjusting the TXP is challenging to reach the desired TXP precisely.

\subsection{Impact of TXP Adjustment on RSSI, Throughput and System Power Consumption Across Three Environments}

\begin{figure*}[htb]
    \centering
    \includegraphics[width=1\textwidth]{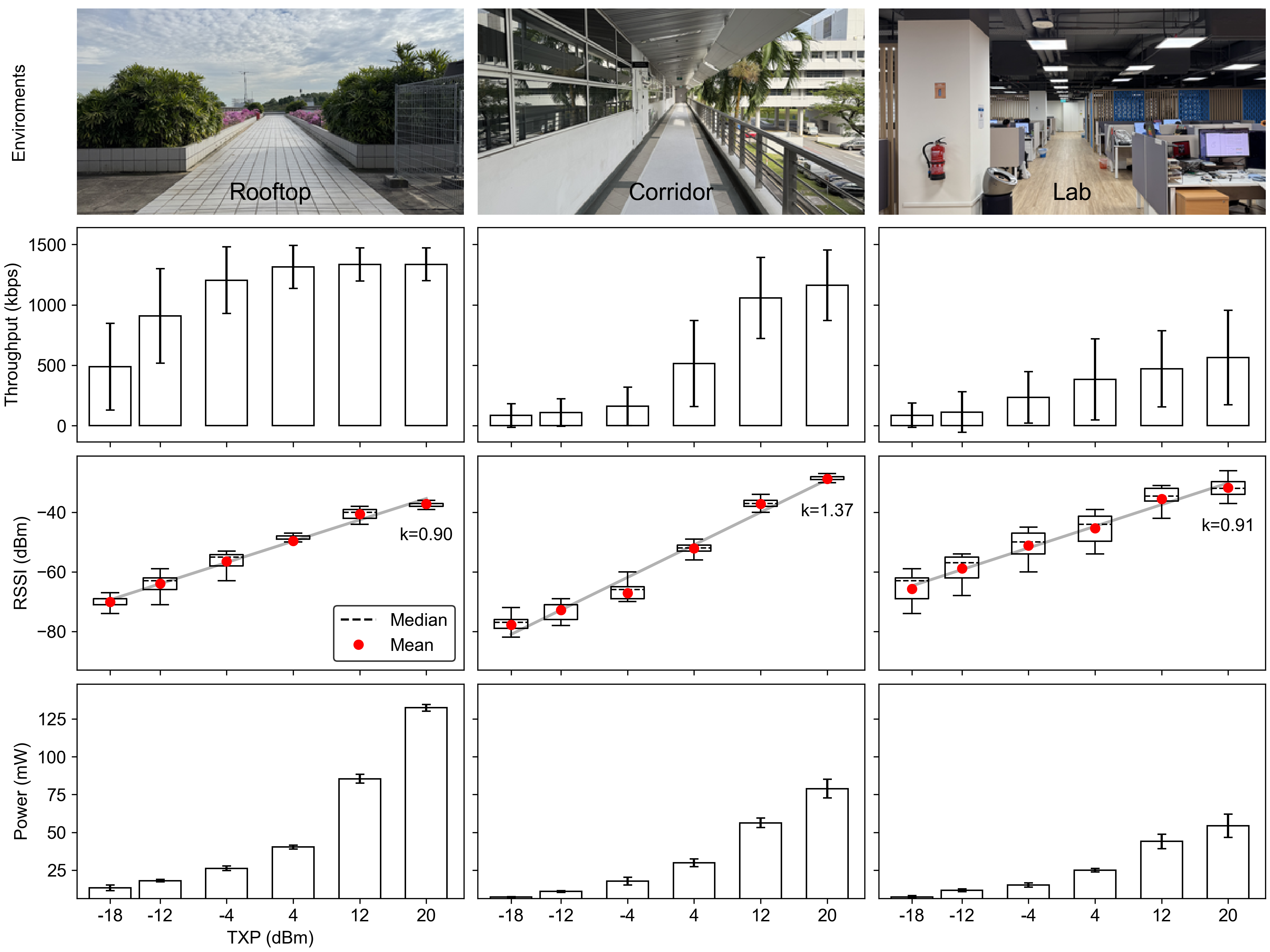}
    \caption{Impact of TXP adjustment on RSSI, throughput, and system power consumption across three environments: evaluating the proportional relationships and environment-specific variations.}
    \label{fig:txp_rssi_throughput}
\end{figure*}

We adjusted the TXP in three different environments to evaluate whether modifying TXP can effectively alter the corresponding RSSI and throughput, as well as to assess the peripheral system power consumption under different TXP levels. The experiment selected TXP levels of -18, -12, -4, 4, 12, and 20 dBm, which, based on previous results, are the effective settings supported by the chipset.
 The experimental results are presented in Fig.~\ref{fig:txp_rssi_throughput}.

In general, higher TXP levels are associated with increased throughput, indicating a proportional relationship between the two. However, the impact of TXP adjustment on throughput varies across different environments. In the rooftop scenario, relatively high throughput can already be achieved at low TXP levels (TXP = -4 dBm can achieve 1200 kbps), and the overall trend does not resemble a linear function. In contrast, for both the corridor and lab environments, the throughput increases approximately linearly with increasing TXP. It is noteworthy that in the rooftop environment, both 12 dBm and 20 dBm TXP levels achieve the maximum throughput. This suggests that the maximum TXP is not necessarily required to attain satisfactory throughput.

Similarly, RSSI exhibits a proportional relationship with TXP. In the rooftop, corridor, and lab environments, the corresponding slopes ($k$) are approximately 0.90, 1.37, and 0.91, respectively—all close to unity. This indicates that for every 1 dB increase in TXP, the RSSI increases by roughly 1 dB, demonstrating the effectiveness of adjusting TXP to control RSSI.

TXP and the peripheral device’s power consumption also exhibit a proportional relationship. However, the rate of power increase varies across the three environments. 

\subsection{Relationship Between Throughput and Overall System Power Consumption}

We selected the throughput data from the previous TXP adjustment lab experiment and compared it with the peripheral system power consumption recorded by PPK. We use the TXP = 20 dBm point as a reference to analyze how variation in throughput affects the system power consumption. This variation is attributed to the decline in signal strength, while the transmitter continues to send data without interruption.

\begin{figure}[t]
 \centering
\includegraphics[width=1\linewidth]{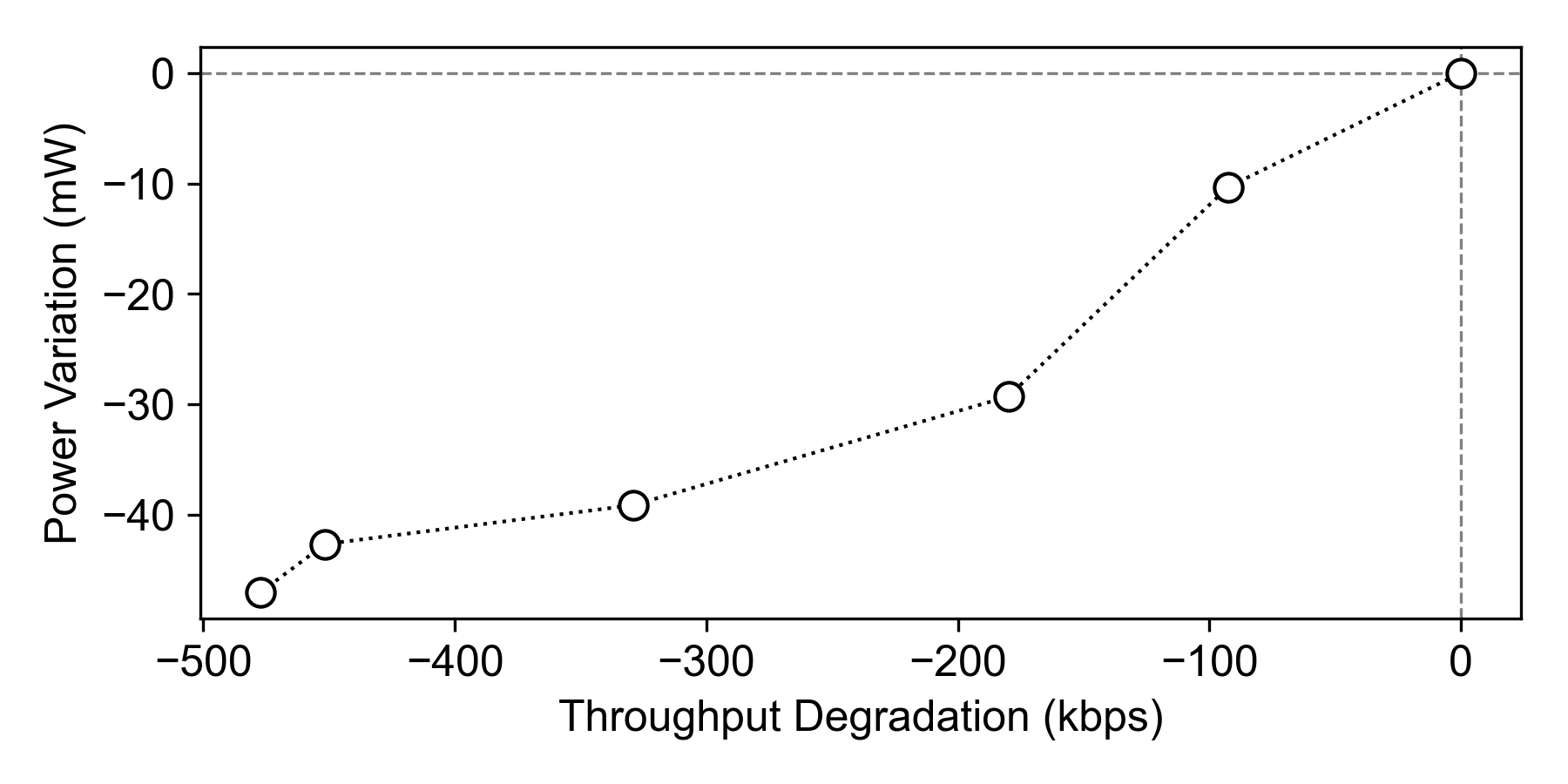}
\caption{Relationship between throughput and overall system power consumption change in the lab scenario.}
\label{fig:throughput_power_corridor}
\end{figure}

The results, as shown in Fig.~\ref{fig:throughput_power_corridor}, indicate a strong correlation between the peripheral device’s system power consumption and throughput. A decrease in throughput is associated with decreased energy consumption. This is primarily due to packet loss. In BLE communication, when the MD bit (More Data) in the LL data-channel PDU header is sent but received with an incorrect CRC, retransmission is required \cite{BluetoothSIG2024CoreV6}. However, retransmission must wait until the next connection event, introducing idle time and reducing radio activity. As a result, less energy is consumed, leading to lower power consumption under poor throughput conditions.

\subsection{Latency from Central to Peripheral Device}

\begin{figure}[t]
    \centering
    \includegraphics[width=1\linewidth]{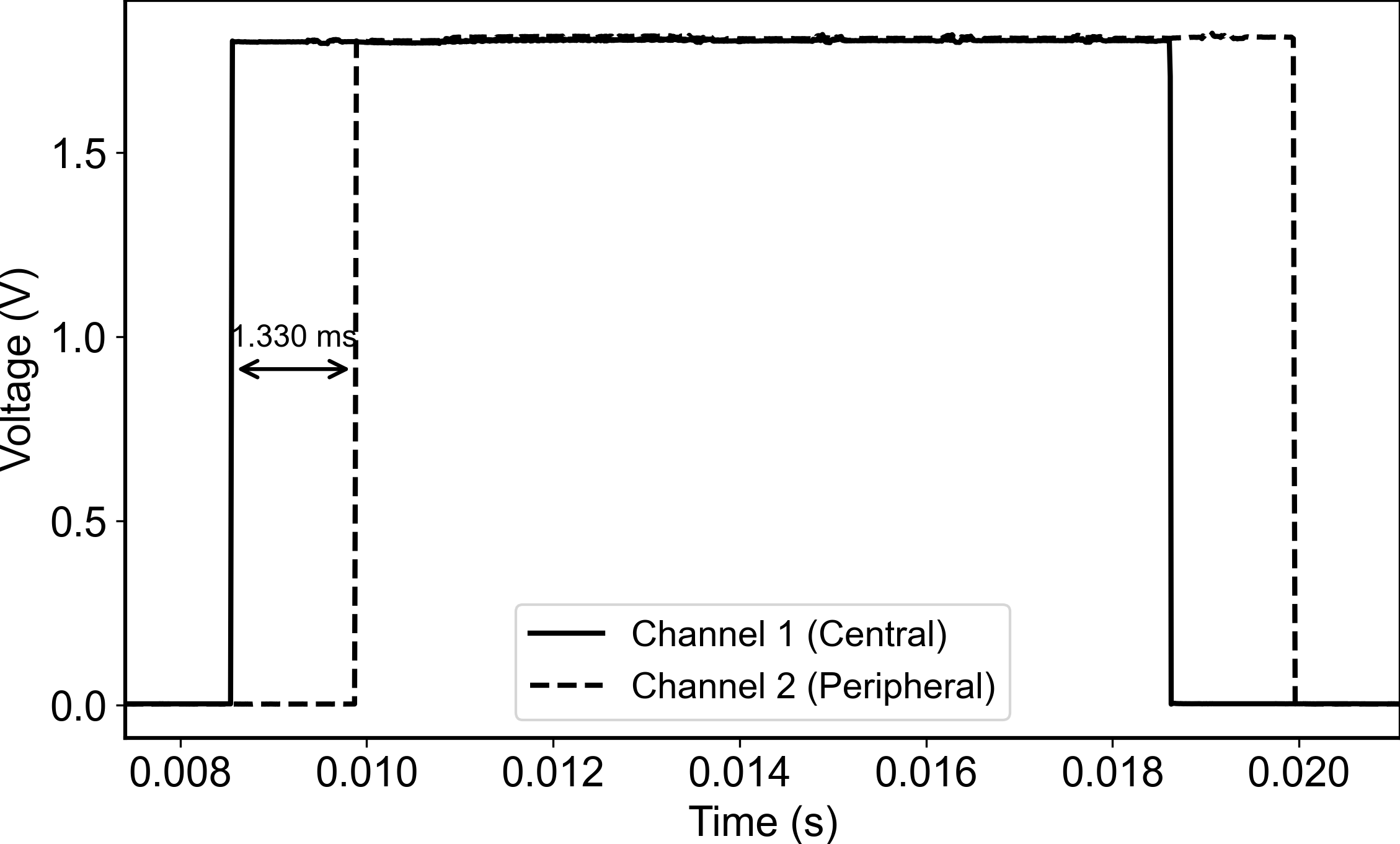}
    \caption{Oscilloscope-based measurement of transmission and update latency between central and peripheral devices in one single test.}
    \label{fig:latency_test}
\end{figure}

By continuously monitoring multiple transmission operations from the central device and the corresponding update operations on the peripheral device using an oscilloscope for 10 times, we measured an average latency of $1.75 \pm 1.02$ ms, which is similar to \cite{s120911734}. One of the single tests is shown in Fig.~\ref{fig:latency_test}. This indicates that the closed-loop control system is capable of rapidly adjusting the peripheral device’s TXP with minimal delay, ensuring timely responsiveness to dynamic conditions. Meanwhile, the associated latency indicates that employing an excessively high PID update frequency yields little benefit.

\section{Results of PID Controller Deployment}

\subsection{RSSI-based Result}

In rooftop environments, any RSSI above -60 dBm consistently yields throughput exceeding 1000 kbps. Therefore, an RSSI-based approach is particularly meaningful under these conditions.

\begin{figure}[t]
    \centering
    \includegraphics[width=1\linewidth]{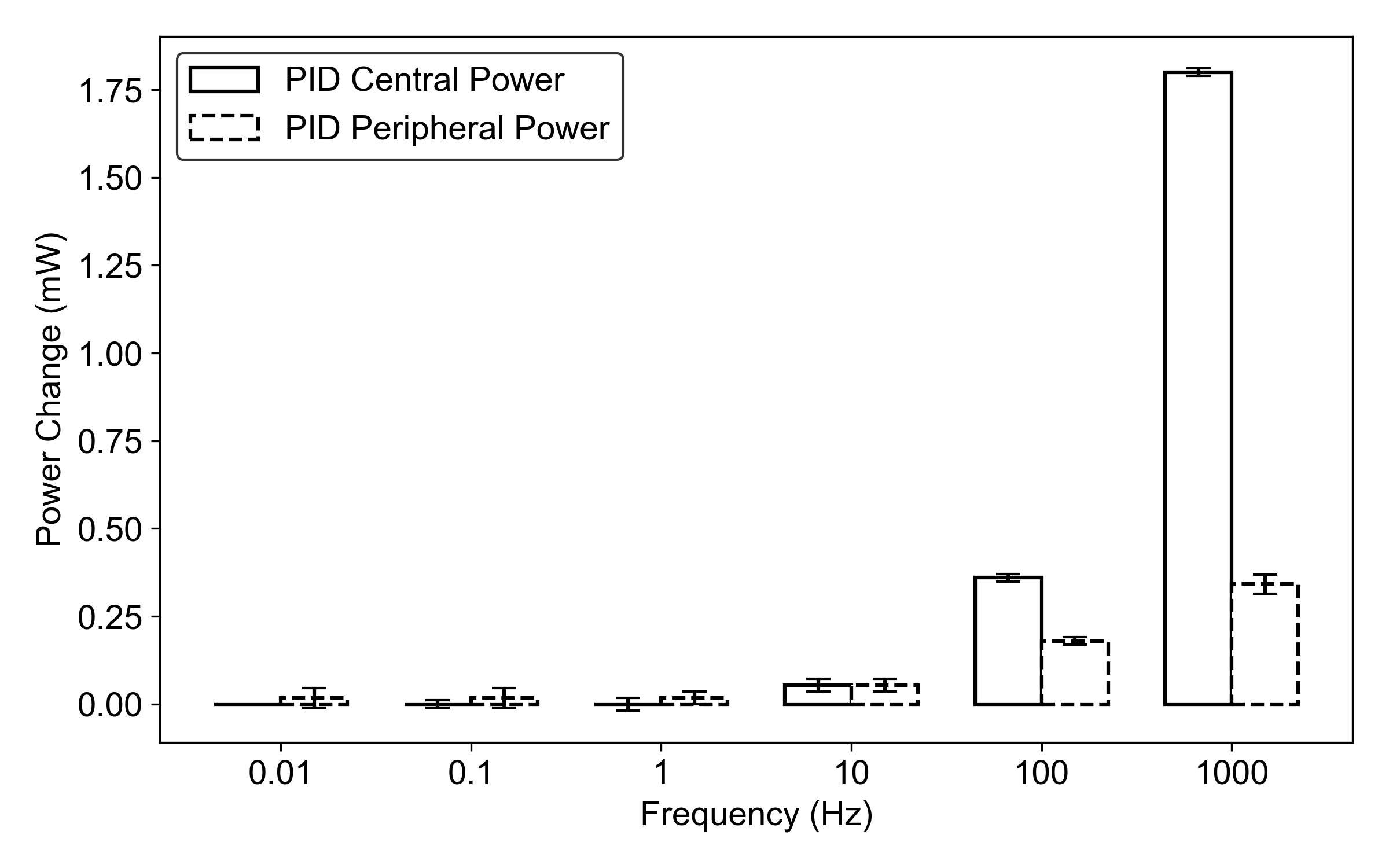}
    \caption{Additional power overhead from RSSI-based TXP control.}
    \label{fig:rssipower_vs_frequency}
\end{figure}

As shown in Fig.~\ref{fig:rssipower_vs_frequency}, we evaluated the additional power consumption induced by RSSI-based PID computation on the central device and TXP updates on the peripheral device across frequencies ranging from 0.01 Hz to 1000 Hz. 

The results indicate that power consumption increases with frequency on both the central and peripheral devices. However, the central device exhibits a much steeper rise in power usage, primarily due to its dominant computational load, which also justifies our decision to implement the controller on the central side to save energy. For the peripheral device, the overall power consumption remains unchanged, mainly between 0.01 Hz and 10 Hz, with only modest increases observed at 100 Hz (0.18 mW) and 1000 Hz (0.34 mW). These results demonstrate that the RSSI-based PID algorithm introduces negligible power overhead on the peripheral side.

\begin{figure}[t]
    \centering
    \includegraphics[width=1\linewidth]{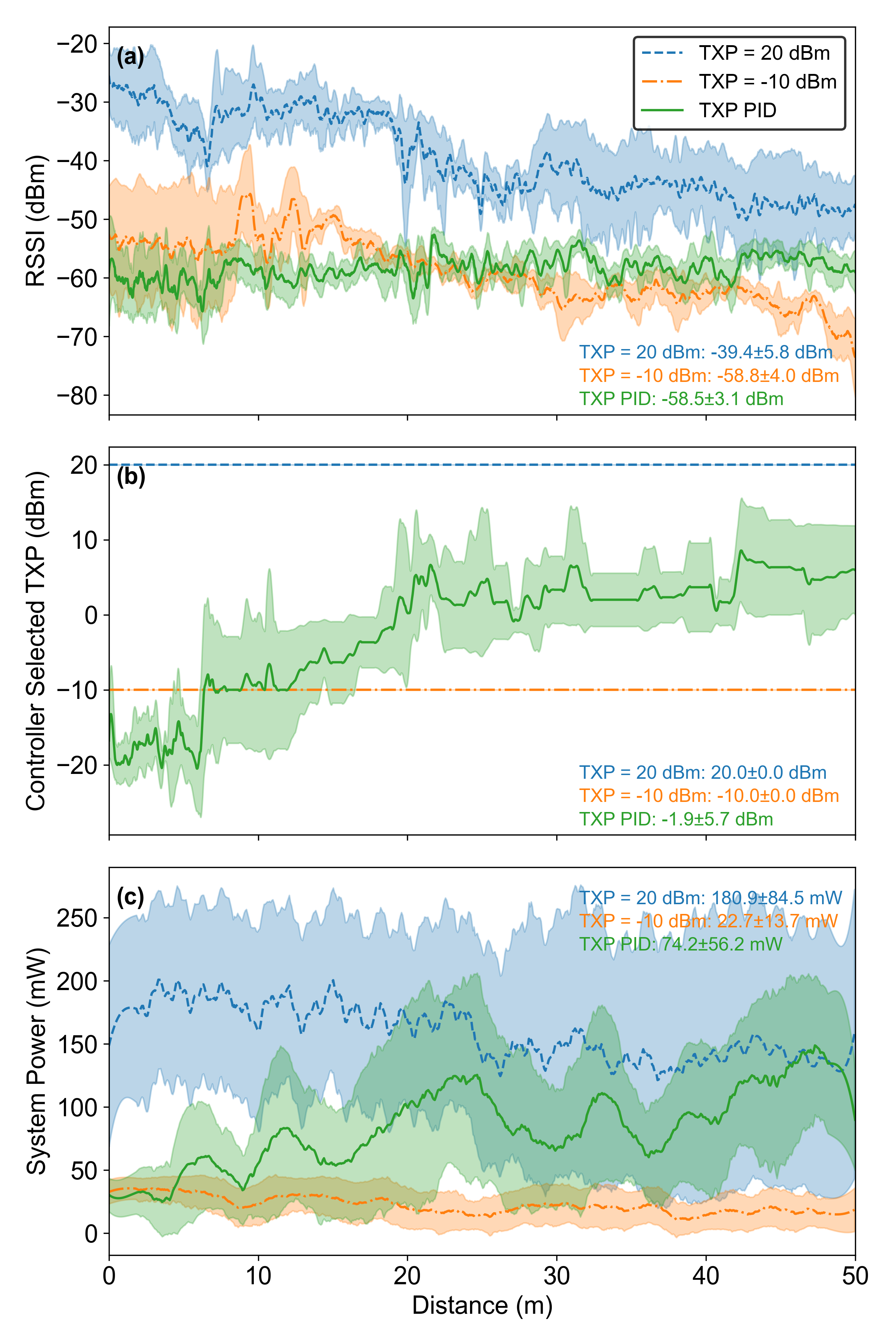}
    \caption{RSSI-based PID control performance in a clean rooftop environment across 0 to 50 m: (a) RSSI variations, (b) controller-regulated TXP, and (c) peripheral device’s system power.}
    \label{fig:rssi_vs_distance_rooftop}
\end{figure}

Considering the insensitivity of RSSI to calculation frequency as shown in Fig.~\ref{fig:sampling_freq_impact}, and taking into account the power consumption at different frequencies illustrated in Fig.~\ref{fig:rssipower_vs_frequency}, we selected 100 Hz as the computation frequency for the PID controller. On the rooftop, the resulting variations of RSSI, controller selected TXP, and peripheral device’s system power with respect to distance are shown in Fig.~\ref{fig:rssi_vs_distance_rooftop}.

For Fig.~\ref{fig:rssi_vs_distance_rooftop}(a), when analyzing the variation of RSSI with distance under fixed TXP conditions of 20 dBm and –10 dBm, the RSSI shows a clear decreasing trend as the distance increases, primarily due to path loss. In contrast, with the PID-controlled system, the green curve maintains the RSSI around the target value of –60 dBm (average: –58.5 dBm), demonstrating the effectiveness of dynamic TXP adjustment. Moreover, PID control improves RSSI stability: compared to standard deviations of 5.8 dB for TXP = 20 dBm and 4.0 dB for TXP = –10 dBm, the TXP PID scheme achieves a lower standard deviation of only 3.1 dB.

Regarding TXP selection, Fig.~\ref{fig:rssi_vs_distance_rooftop}(b) shows that the TXP chosen by the PID controller increases with distance. This trend indicates that higher TXP is applied to compensate for the path loss introduced by increasing distance. Under the 50-meter experimental setting, the average TXP utilized by the PID controller is -1.9 dBm, which falls between the fixed TXP settings of -10 dBm and 20 dBm.

For the peripheral-side system power consumption shown in Fig.~\ref{fig:rssi_vs_distance_rooftop}(c), a decreasing trend is observed when using fixed TXP values. This is consistent with the behavior described in Fig.~\ref{fig:throughput_power_corridor}, where power consumption drops due to increased packet loss. In contrast, with the PID-based method, overall system power consumption increases as the controller dynamically raises TXP to compensate for path loss caused by increasing distance. Overall, the RSSI-based method reduces power consumption by 60\% compared to the fixed TXP = 20 dBm configuration, while still achieving satisfactory throughput under rooftop clean environmental conditions.

\begin{figure}[t]
    \centering
    \includegraphics[width=1\linewidth]{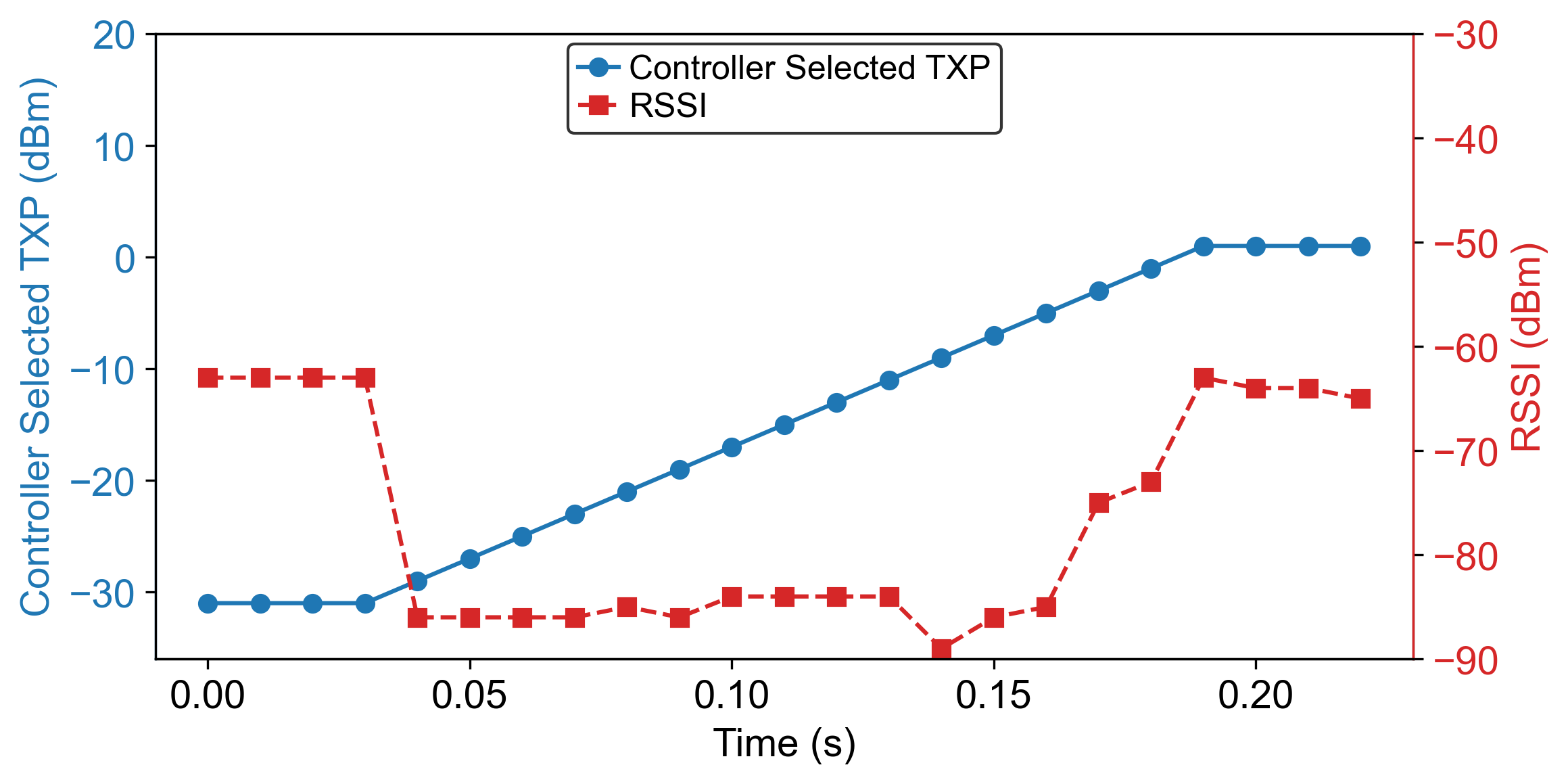}
    \caption{System response of RSSI-based method under extreme conditions. Variation in RSSI and TXP. }
    \label{fig:extreme_rssi_response}
\end{figure}

We also evaluated the stability of the RSSI-based method under extreme conditions by modifying the target RSSI to –65 dBm in the lab environment and artificially inducing signal degradation. This was achieved by removing the FEM on the receiver side, causing the RSSI to approach the disconnection threshold. As shown in Fig.~\ref{fig:extreme_rssi_response}, the removal of the FEM led to an abrupt RSSI drop of nearly 20 dB. In response, the controller immediately increased the TXP to counteract the loss, and the RSSI returned to the target level of approximately –65 dBm within about 150 ms. The blue controller selected TXP increases in a highly linear manner because the incremental PID output is limited to 2 dB per step to prevent overshooting.

Overall, the RSSI-based method performs effectively in clean environments, offering stable RSSI and high control frequency, while also demonstrating reliable operation under extreme conditions. However, its lack of direct correlation with throughput limits its applicability across broader scenarios.

\subsection{Throughput-based Result}

The corridor presents a more complex propagation environment, where the relationship between throughput and RSSI is less predictable, as indicated by Fig.~\ref{fig:rssi_throughput_env}. Thus, maintaining RSSI is less meaningful, necessitating the use of a throughput-based control strategy.

\begin{figure}[t]
    \centering
    \includegraphics[width=1\linewidth]{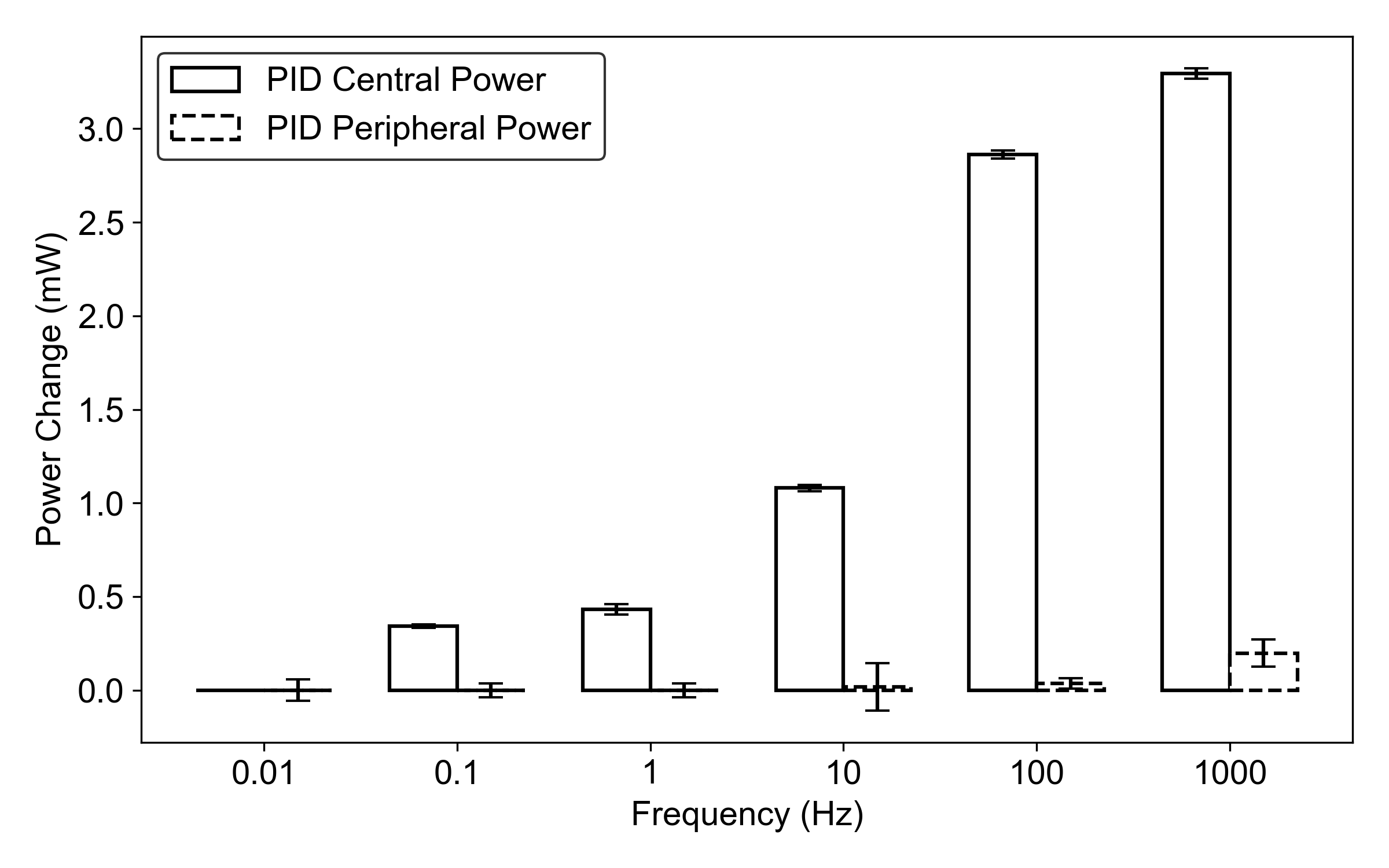}
    \caption{Additional power overhead from throughput-based TXP control.}
    \label{fig:throughput_sampling_freq}
\end{figure}

The additional power consumption introduced by throughput-based PID computation on the central device and TXP updates on the peripheral device at different frequencies is shown in Fig.~\ref{fig:throughput_sampling_freq}. Overall, higher frequencies result in greater power consumption. Notably, the computational overhead on the central device is significantly higher than that of the RSSI-based method: for example, at a computation frequency of 1000 Hz, the power consumption of the throughput-based method calculated at the central end is approximately twice that of the RSSI-based approach, as calculating throughput requires substantially more processing resources than reading RSSI values. In contrast, the peripheral device exhibits a similar trend to that observed in the RSSI-based approach. Based on these findings and in consideration of Fig.~\ref{fig:sampling_freq_impact}, to balance throughput measurement stability with responsiveness, we selected 1 Hz as the PID frequency for our throughput-based PID control, and the experiment was conducted in the corridor.

\begin{figure}[t]
    \centering
    \includegraphics[width=1\linewidth]{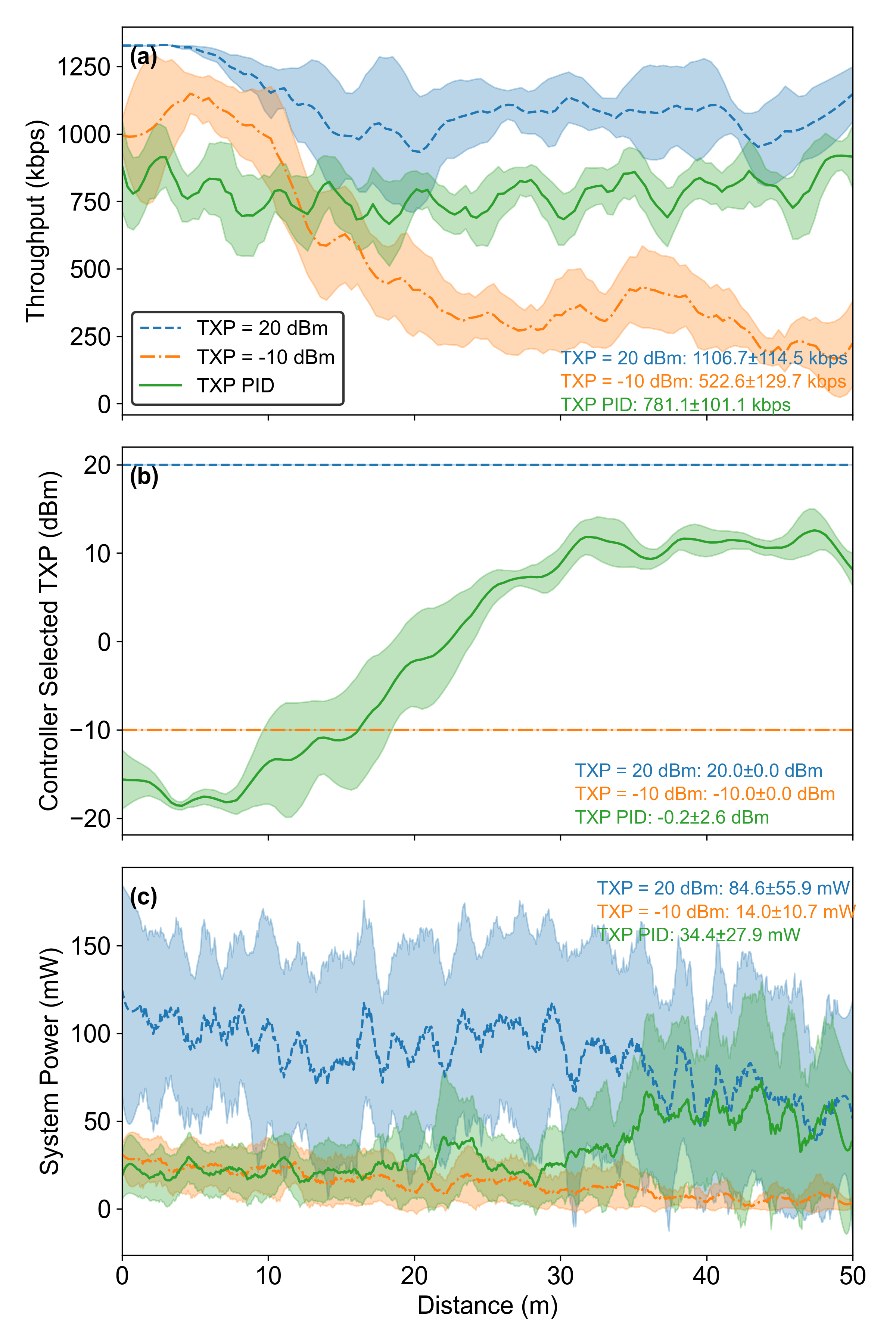}
    \caption{Throughput-based PID control performance in a complex corridor environment across 0 to 50 m: variations of (a) throughput, (b) controller-regulated TXP, and (c) peripheral device’s system power.}
    \label{fig:throughput_range}
\end{figure}

In the distance-varying experiment ranging from 0 to 50 m, the results are summarized in Fig.~\ref{fig:throughput_range}. For fixed TXP strategies shown in Fig.~\ref{fig:throughput_range}(a), throughput consistently declined as distance increased. In contrast, the PID-based method effectively maintained throughput near the 800 kbps target (average: 781.1 kbps) while also reducing its standard deviation (101.1 kbps), compared to 114.5 kbps for TXP = 20 dBm and 129.7 kbps for TXP = –10 dBm. 

Regarding TXP behavior shown in Fig.~\ref{fig:throughput_range}(b), the PID controller adaptively increased TXP with distance, following a trend similar to the RSSI-based strategy, and selected an average TXP of –0.2 dBm. In terms of peripheral power consumption shown in Fig.~\ref{fig:throughput_range}(c), both fixed-TXP strategies exhibited a decreasing trend with a more extended range, similar to Fig.~\ref{fig:throughput_power_corridor}. However, the PID-controlled system consistently sustained power consumption around 34.4 mW, representing a 60\% reduction compared to using a fixed TXP of 20 dBm. This suggests that the energy required to maintain the target throughput remained nearly constant across all distances.

\begin{figure}[t]
    \centering
    \includegraphics[width=1\linewidth]{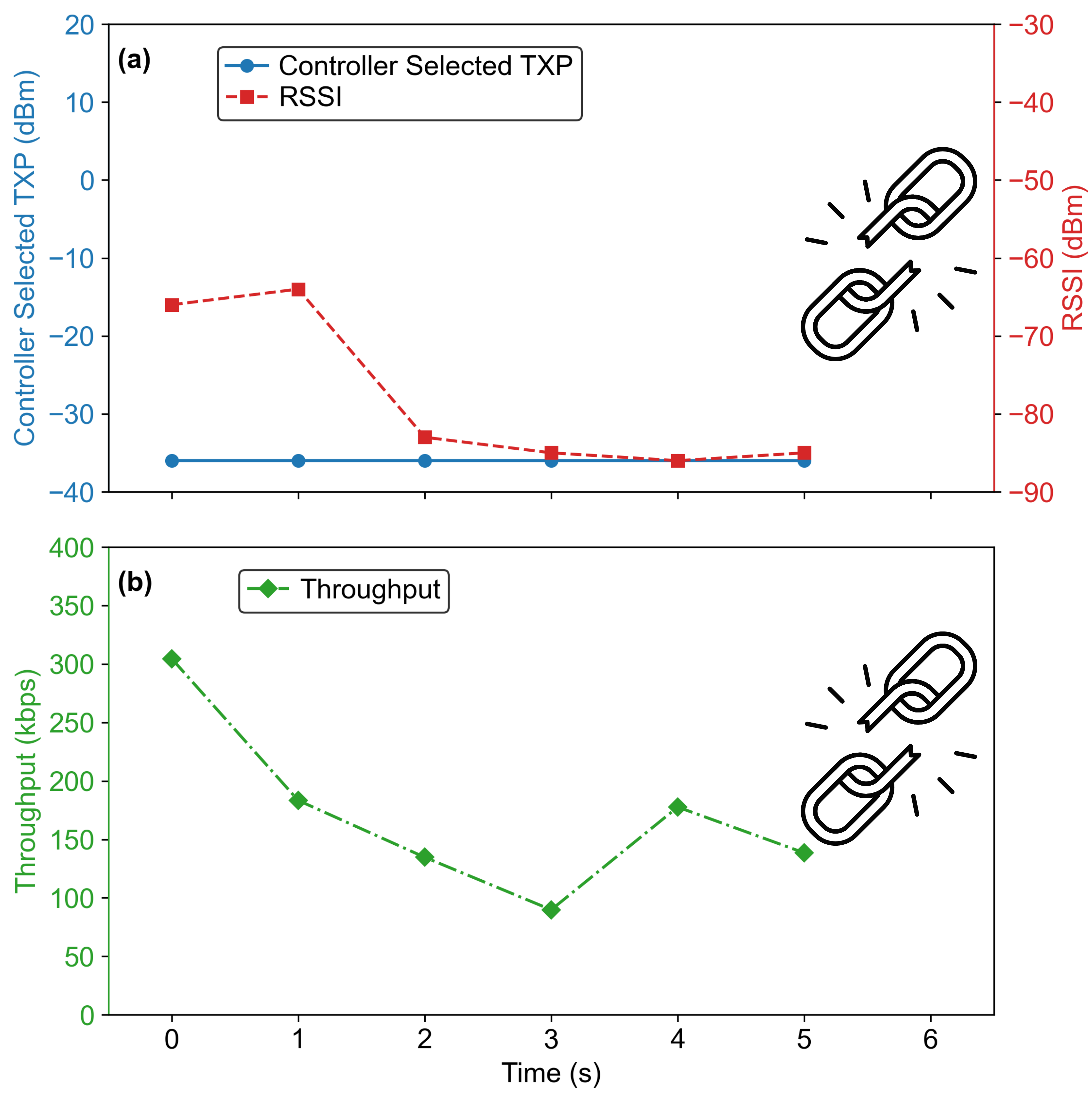}
    \caption{System response of throughput-based method under extreme conditions. (a) Variation of TXP and RSSI. (b) Variation of throughput.
}
    \label{fig:extreme_throughput_failure}
\end{figure}

To evaluate the effectiveness of the throughput-based method under extreme conditions, we lowered the target throughput to 100 kbps and set the distance between the peripheral and central devices to 30 cm in the lab, resulting in an RSSI of –65 dBm. Following the same approach, we removed the FEM to simulate an extreme scenario. The experimental results are shown in Fig.~\ref{fig:extreme_throughput_failure}. At the end of the first second, the RSSI dropped sharply due to FEM removal as shown in Fig.~\ref{fig:extreme_throughput_failure}(a). However, due to the inherent latency and fluctuations in throughput feedback shown in Fig.~\ref{fig:extreme_throughput_failure}(b), the system maintained communication for 3 seconds before disconnecting.

Overall, the throughput-based method enables intuitive control of stable throughput even in complex environments; however, due to its slower control response and the inherent fluctuation of throughput, it fails to ensure reliable communication under extreme conditions, such as plug-out.

\subsection{RSSI+Througput Result}

In our final controller design, the throughput loop updates the RSSI target at 1 Hz, while the RSSI loop runs at 100 Hz to stabilize the RSSI around the computed target. This configuration results in an additional power consumption of 0.19 mW on the peripheral side due to frequent TXP updates. On the central side, the dual-loop operation and associated data transmission increase power consumption by 2.39 mW. Notably, this total does not equal the simple sum of standalone 1 Hz throughput and 100 Hz RSSI loops, likely due to low-power sleep behavior in the Zephyr environment—activating more tasks reduces MCU sleep opportunities, thereby elevating overall system power consumption \cite{zephyrproject2025docs}.

\begin{figure}[t]
    \centering
    \includegraphics[width=1\linewidth]{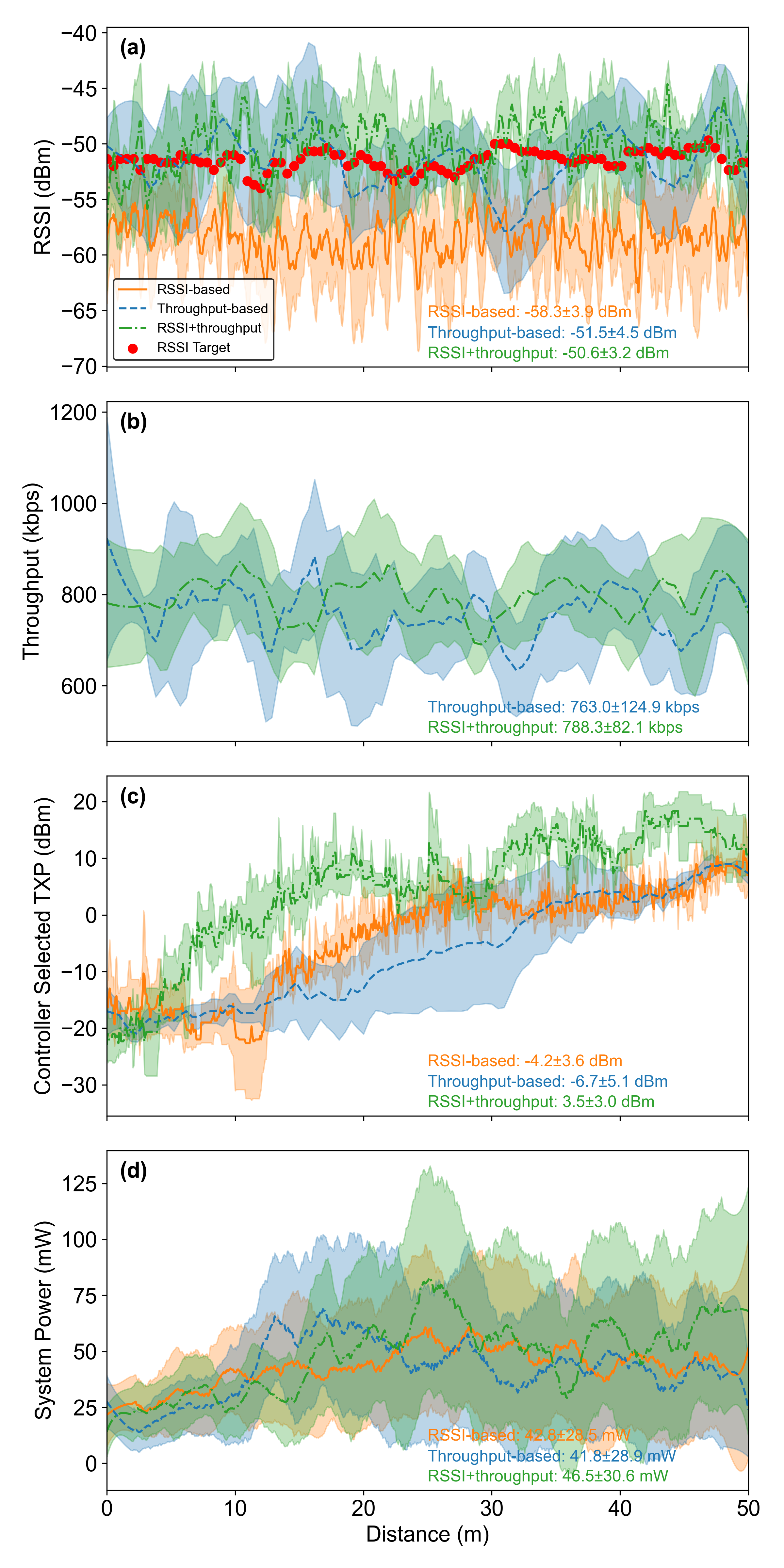}

    \caption{Comparison between the RSSI-based, throughput-based, and hybrid RSSI + throughput methods across 0 to 50 m: (a) RSSI, (b) throughput, (c) controller-regulated TXP, and (d) peripheral device’s system power.}

    \label{fig:dual_loop_range}
\end{figure}

In a complex lab environment, we evaluated three approaches—an RSSI-based method, a throughput-based method, and an RSSI+throughput hybrid method—as illustrated in Fig.~\ref{fig:dual_loop_range}. Specifically, we compared (a) the RSSI measurements, (b) the data throughput results, (c) the TXP adjustments enacted by each method, and (d) the corresponding power consumption of the peripheral device under each scheme.

In Fig.~\ref{fig:dual_loop_range}(a), the RSSI-based method exhibits signal oscillations around the –60 dBm target, with an average value of –58.3 dBm. Under the throughput-based method, although RSSI is not directly regulated, it fluctuates with an average of -51.5 dBm. In this scenario, the strong correlation between RSSI and throughput in the given environment enables the system to adjust the TXP to achieve the 800 kbps throughput target, thereby driving the RSSI into a specific operational range. Our hybrid RSSI+throughput approach uses the throughput-loop to dynamically update the RSSI target, after which the RSSI-loop maintains RSSI around this new target. The average RSSI for the hybrid method was measured to be -50.6 dBm. The sparsity observed in both the throughput-based curve and the RSSI target under the RSSI+throughput method is due to their shared 1 Hz update rate.

In Fig.~\ref{fig:dual_loop_range}(b), due to the limitations of the RSSI-based method, no throughput readings are available; therefore, only the throughput-based and the RSSI+throughput approaches yield data. Both methods stabilize throughput around 800 kbps, but our RSSI+throughput method achieves a value closer to the target (788.3 kbps vs. 763.0 kbps) with lower standard deviation and reduced fluctuation (82.1 kbps vs. 124.9 kbps). This improvement likely stems from using RSSI as the TXP control variable, which helps mitigate the variability inherent in throughput measurements while enhancing the system’s responsiveness.

In Fig.~\ref{fig:dual_loop_range}(c), the selected TXP for all three methods increases with distance. The RSSI-based and RSSI+throughput approaches exhibit higher adjustment density than the throughput-based method, as the throughput method updates at 1 Hz while the former two operate at 100 Hz. Across the entire 50 m range, the average TXP selected by the RSSI-based method is –4.2 dBm, by the throughput-based method is –6.7 dBm, and by the RSSI+throughput method is 3.5 dBm.

In Fig.~\ref{fig:dual_loop_range}(d), the power consumption across all methods remains at a comparable level, validating that the hybrid method can also achieve substantial power savings compared to fixed maximum TXP settings. For the RSSI-based, throughput-based, and RSSI+throughput methods, the average power consumption is 42.8 mW, 41.8 mW, and 46.5 mW, respectively. These values are proportional to the TXP levels shown in Fig.~\ref{fig:dual_loop_range}(c).

\begin{figure}[t]
    \centering
    \includegraphics[width=1\linewidth]{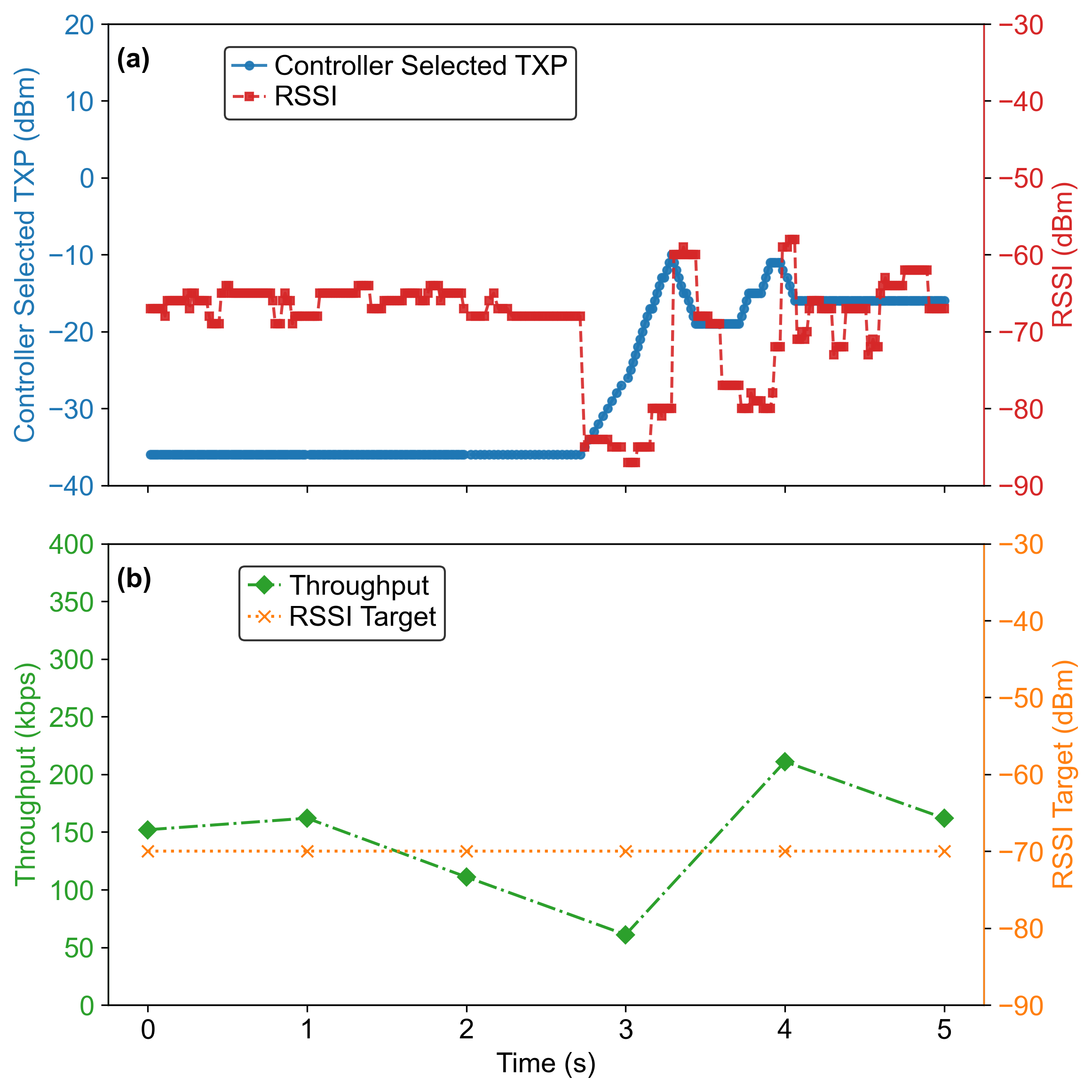}
    \caption{System response of RSSI+throughput  method under extreme conditions. (a) Variation in RSSI and TXP within the RSSI loop. (b) Variation in throughput and RSSI target within the throughput loop.
}
    \label{fig:dual_loop_extreme}
\end{figure}

In the extreme-case test, the RSSI-based method effectively resisted sudden interference. It returned RSSI to the target value as shown in Fig.~\ref{fig:extreme_rssi_response}, whereas the throughput-based method resulted in a disconnection due to its slower response as demonstrated in Fig.~\ref{fig:extreme_throughput_failure}. To combine the advantages of both approaches, we propose the hybrid RSSI+throughput method. Similar to the previous experiment, we target a throughput of approximately 100 kbps and set the distance between the peripheral and central devices to 30 cm in the lab.

As shown in Fig.~\ref{fig:dual_loop_extreme}(a), at the end of the third second, the FEM was removed, causing a significant 20 dB drop in RSSI. The RSSI loop rapidly increased the TXP to counteract this change. During this period, throughput fluctuated around the 100 kbps target as shown in Fig.~\ref{fig:dual_loop_extreme}(b), and the peripheral device remained connected. 

Overall, the RSSI–throughput hybrid method maintains the required throughput in complex environments, as the fast response of the RSSI loop compensates for the delay and fluctuation of the throughput loop, and it effectively prevents disconnections under sudden perturbations.

\section{Conclusion and Future Work}

This study systematically investigated the interplay between RSSI, throughput, TXP, and system power consumption in BLE communication under various environmental conditions. Based on this analysis, we designed and implemented a real-time closed-loop TXP control framework using PID controllers. The system supports three modes: RSSI-based, throughput-based, and a cascaded hybrid RSSI+throughput method.

Experimental results demonstrate that:
\begin{itemize}
  \item The RSSI-based method enables rapid responsiveness, making it suitable for clean or fast-changing environments. It is also ideal for ultra-low throughput applications, where maintaining a stable connection suffices.
  \item The throughput-based method directly regulates throughput in complex environments but exhibits a delayed response under abrupt perturbations. It is better suited for gradually changing scenarios without sudden disruptions.

  \item The proposed RSSI+throughput hybrid strategy outperforms both individual methods by maintaining target-level throughput with small standard deviation in dynamic, complex environments while reducing power consumption and preventing disconnection during sudden signal degradation.

\end{itemize}

The dual-loop controller runs efficiently on the central device with minimal peripheral-side overhead, making it well-suited for real-time BLE applications.

In future work, we plan to extend this framework to BLE Mesh networks and heterogeneous multi-node systems, where global coordination of TXP becomes essential. We will also explore adaptive PID parameter tuning using machine learning to improve responsiveness under mobility and traffic variation. Moreover, validating this control framework on commercial BLE SoCs across vendors will be crucial for practical deployment in consumer-grade IoT systems.

\bibliographystyle{IEEEtran}
\bibliography{main}

 \vspace{-43pt}


\begin{IEEEbiography}[{\includegraphics[width=1in,height=1.25in,clip,keepaspectratio]{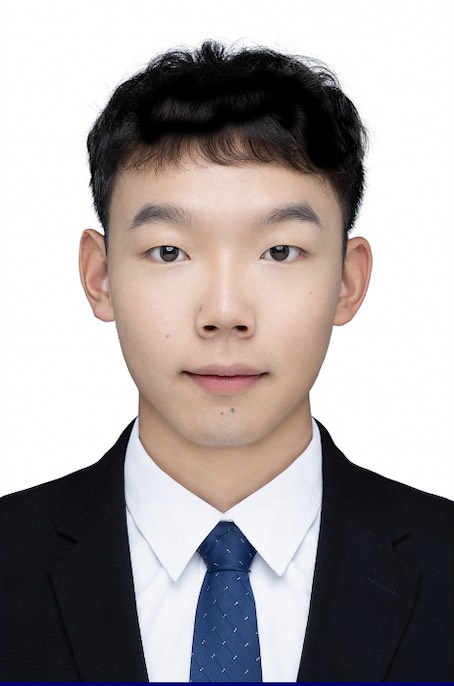}}]{Ziyao Zhou} received the B.Eng. degree in Communication Engineering from the University of Electronic Science and Technology of China, Chengdu, China, in 2023, and the M.Sc. degree in Communication Engineering from Nanyang Technological University, Singapore, in 2024. He is currently working toward a Ph.D. in the Digital, AI, Robotics, and Electronics (DARE) Lab for Translational Medicine, School of Electrical and Electronic Engineering, Nanyang Technological University. His research interests include wireless sensor and actuator networks for gastrointestinal tract applications.
\end{IEEEbiography}

 \vspace{-25pt}

\begin{IEEEbiography}[{\includegraphics[width=1in,height=1.25in, clip,keepaspectratio]{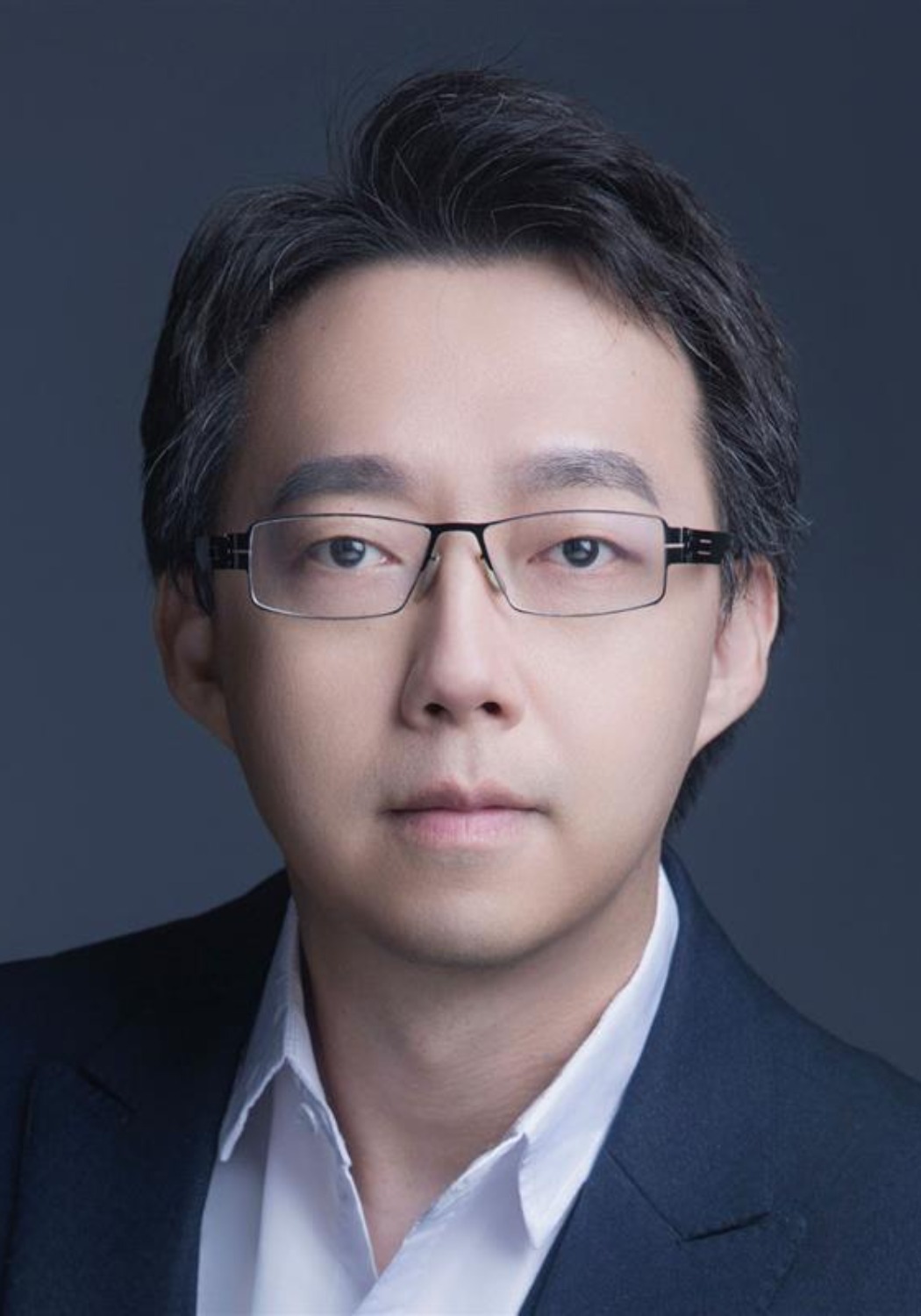}}]{Hen-Wei Huang} received the B.S. and M.S. degrees in mechanical engineering from the National Taiwan University, Taiwan, in 2011 and 2012, respectively, and the Ph.D. degree in robotics from ETH Zürich in 2018.

He is currently an Assistant Professor at the School of Electrical and Electronic Engineering as well as the LKC School of Medicine, Nanyang Technological University, Singapore. He is also the director of the DARE Lab. His research interests include in vivo wireless sensor networks, ingestible electronics, robotic-assisted drug delivery, and translational medicine.
\end{IEEEbiography}

%


\end{document}